\documentclass{aa}  

\usepackage{natbib}
\bibpunct{(}{)}{;}{a}{}{,} 
\usepackage{graphics}   
\usepackage{graphicx}   
\usepackage{subfig}
\usepackage{epstopdf}
\usepackage{amsmath} 
\usepackage[varg]{txfonts}
\usepackage{pdflscape}
\usepackage{hyperref}   
\usepackage{longtable}
\usepackage{balance}
\usepackage{color}
\usepackage[normalem]{ulem}

\hypersetup{
    bookmarks=true,         
    colorlinks=true,       
    linkcolor=blue,          
    citecolor=blue,        
    filecolor=blue,      
    urlcolor=blue           
}

\def\ga{\,\,\raise0.14em\hbox{$>$}\kern-0.76em\lower0.28em\hbox{$\sim$}\,\,}
\def\la{\,\,\raise0.14em\hbox{$<$}\kern-0.76em\lower0.28em\hbox{$\sim$}\,\,}

\def\Msun{$M_{\odot}$}

\def\cm3{cm$^{-3}$}

\def\chem#1#2{$\mathrm{^{#2}\kern-0.8pt#1}$}
\def\reac#1#2#3#4#5#6{$\mathrm{\, ^{#2}\kern-0.8pt{#1}\, ({#3}\, ,{#4})\, {}^{#6}\kern-0.8pt{#5}\, }$}

\def\env{\mathrm{env}} 
\def\top{\mathrm{top}} 
\def\bot{\mathrm{bot}}

\def\be{\begin{equation}} 
\def\ee{\end{equation}}
\def\beqy{\begin{eqnarray}}
\def\eeqy{\end{eqnarray}}
\def\bmlet{\begin{mathletters}}
\def\emlet{\end{mathletters}}

\begin{document}

\title{The intermediate neutron capture process}
\subtitle{V. The i-process in AGB stars with overshoot}

\author{A. Choplin   
\and 
L. Siess
\and
S. Goriely
\and
S. Martinet}
\offprints{arthur.choplin@ulb.be}

\institute{
Institut d'Astronomie et d'Astrophysique, Universit\'e Libre de Bruxelles,  CP 226, B-1050 Brussels, Belgium
}

\date{Received --; accepted --}

\abstract
{
The intermediate neutron capture process (i-process) can develop during proton ingestion events (PIE), potentially during the early stages of low-mass low-metallicity asymptotic giant branch (AGB) stars.
}
{
We examine the impact of overshoot mixing on the triggering and development of i-process nucleosynthesis in AGB stars of various initial masses and metallicities.
}
{
We computed AGB stellar models, with initial masses of 1, 2, 3, and 4~\Msun{} and metallicities in the $-2.5 \le $~[Fe/H]~$\le 0$ range, using the stellar evolution code \textsf{STAREVOL} with a network of 1160 nuclei coupled to the transport equations. We considered different overshooting profiles below and above the thermal pulses, and below the convective envelope. 
}
{
The occurrence of PIEs is found to be primarily governed by the amount of overshooting at the top of pulse  ($f_{\top}$) and to increase with rising $f_{\top}$. 
For $f_{\top} = 0$, 0.02, 0.04, and 0.1, we find that 0~\%, 6~\%, 24~\%, and 86~\% of our 21 AGB models with $-2<$~[Fe/H]~$<0$ experience a PIE, respectively.  Variations of the overshooting parameters during a PIE leads to a scatter on abundances of $0.5 - 1$~dex on elements, with $36<Z<56;$  however, this barely impacts the production of elements with $56<Z<80$, which therefore appear to be a reliable prediction of our models. Actinides are only produced if the overshooting at the top of pulse is small enough. 
We also find that PIEs leave a $^{13}$C-pocket at the bottom of the pulse that can give rise to an additional radiative s-process nucleosynthesis. 
In the case of the 2~\Msun{} models with [Fe/H]~$=-1$ and $-0.5$, it produces a noticeable mixed i+s chemical signature at the surface. 
Finally, the chemical abundance patterns of {22 observed r/s-stars candidates (18 dwarfs or giants and 4 post-AGB)} with $-2<$~[Fe/H]~$<-1$ are found to be in reasonable agreement with our AGB model predictions. 
{The binary status of the dwarfs/giants being unclear, we suggest that these stars have acquired their chemical pattern either from the mass transfer of a now-extinct AGB companion or from an early generation AGB star that polluted the natal cloud. } 
}
{
The occurrence of PIEs and the development of i-process nucleosynthesis in AGB stars remains sensitive to the overshooting parametrization. A high (yet realistic) $f_{\top}$ value triggers PIEs at (almost) all metallicities. 
{The existence of r/s-stars at [Fe/H]~$ \simeq -1$ is in favour of an i-process operating in AGB stars up to this metallicity.}
Stricter constraints from multi-dimensional hydrodynamical models on overshoot coefficients could deliver new insights into the contribution of AGB stars to heavy elements in the Universe.  
}


\titlerunning{}

\authorrunning{A. Choplin et al. }

\maketitle


\section{Introduction}
\label{sect:intro}

Understanding the chemical evolution of the Universe requires a thorough comprehension of all nucleosynthetic processes, in particular, the astrophysical sites where they take place and the various physical mechanisms taking place at their origin. 
The elements heavier than Fe are mostly synthesized by the slow and rapid neutron-capture processes \citep[e.g.,][for a review]{arnould20}, which are characterized by neutron densities on the order of $N_n = 10^5 - 10^{10}$ and $N_n>10^{24}$~cm$^{-3}$, respectively. The main astrophysical sites for the s-process are asymptotic giant branch stars (AGB) {\citep[e.g.,][]{gallino98,herwig05,cristallo09a,cristallo09b,cristallo11,bisterzo11,karakas14, fishlock14,goriely18c}} and the helium-burning core of massive stars \citep[e.g.,][]{langer89,prantzos90,pignatari08,frischknecht16,Choplin18}. 
The r-process requires a more violent environment, which could be the merging of two neutron stars \citep[e.g.,][]{arnould07, goriely11a, goriely11b, wanajo14, just15}, collapsars, or magnetorotational supernovae \citep{winteler12,nishimura15,siegel19}.

Other secondary neutron capture processes were identified. One of them is the so-called intermediate neutron-capture process \citep[][]{cowan77}, which is associated with neutron densities of about $N_n = 10^{13} - 10^{16}$~cm$^{-3}$, namely, in between the s- and r-processes. 
The existence of the i-process is supported by the observation of stars with chemical overabundances that are not compatible with the s- or the r-processes alone. Instead these stars can be explained by considering an intermediate neutron irradiation, as modeled by i-process calculations \citep[r/s-stars, e.g.,][]{mishenina15, roederer16, karinkuzhi21, karinkuzhi23, mashonkina23, hansen23}.
Some pre-solar grains may also bear the signature of i-process nucleosynthesis \citep{fujiya13,jadhav13,liu14}. 

The i-process nucleosynthesis can develop if protons are mixed in a convective helium-burning zone (proton ingestion event or PIE). 
This could take place in different astrophysical sites \citep[see][for a detailed list]{choplin21}, one of which is low-metallicity low-mass AGB stars \citep[e.g.,][]{iwamoto04,cristallo09a,suda10,choplin21,choplin22a,goriely21,gilpons22}. 
In such stars, the top of the convective thermal pulse encroaches on the tail of the hydrogen-rich zone. Protons are then transported downwards by convection (on a typical timescale of 1~hr) and burnt on the way via $^{12}$C($p,\gamma$)$^{13}$N. The $^{13}$N isotope decays to $^{13}$C in about 10~min. At this point, the $^{13}$C($\alpha,n$)$^{16}$O reaction becomes active, mostly at the bottom of the pulse, where the temperature is about 250~MK. 
The neutron density goes up to $N_n \simeq 10^{15}$~cm$^{-3}$, which triggers the i-process, and possibly the synthesis of actinides \citep{choplin22b}.
The convective thermal pulse splits in two parts after the neutron density peak \citep[cf. Sect.~3.5 in][for a discussion about the split]{choplin22a} and the upper part eventually merges with the convective envelope. There, the i-process products are diluted into the envelope and then expelled by the stellar winds.

In \citet[hereafter paper III]{choplin22a}, we have shown that AGB models with $M_{\rm ini} \lesssim 3.0$ and [Fe/H]~$\lesssim -2$ that do not include extra mixing (e.g., overshoot) experience PIEs (see Fig. 3 in paper III for a more detailed view). 
What remains unknown is how the development of PIEs in AGB stars is affected by extra mixing, particularly overshooting. 
This is of importance to better assess the contribution of AGB stars to the Galactic enrichment as well as the chemical evolution of the Universe. 

Hydrodynamical simulations have shown that convection extends beyond the boundary of the convectively unstable region \citep[e.g.,][]{freytag96}. 
This extra mixing can be implemented in 1D stellar evolution codes with a parametrized diffusion coefficient. One possible parametrization is expressed as follows \citep[e.g.,][]{freytag96, herwig97}:
\begin{equation}
D_{\rm over}(z) = D_{\rm cb} \exp \left( \frac{-2 \, z}{f_{\rm over} \, H_P} \right)
,\end{equation}
where $z$ is the distance from the formal convective boundary, $D_{\rm cb}$ is the diffusion coefficient at the edge of
the convective zone (as defined by the Schwarzschild criterion), $f_{\rm over}$ is a free parameter controlling the efficiency of the mixing and that can be different at each convective boundary. 
Table~\ref{table:ff} reports the values of $f_{\rm over}$ adopted at three different convective borders by various studies.
We note that \cite{herwig07} proposed to approximate the overshoot at the base of the convective pulse by two exponential decays \citep[see also][]{battino16},
characterized by  $f_{\rm bot, 1} = 0.01$ and $f_{\rm bot, 2} = 0.14$ (hence, the two values in Table~\ref{table:ff}). 
The overshoot parameters are generally calibrated based on observations. 
For instance, the overshoot parameter below the envelope, $f_{\env}$, is determine so as to form a $^{13}$C-pocket massive enough to account for the level of surface s-process enrichment. 
The overshoot parameter at the bottom of the pulse,  $f_{\bot}$, was shown to scale with the O abundance in the thermal pulse \citep{herwig00}. This parameter was thus sometimes calibrated to reproduce the O abundance of observed post-AGB stars \citep[e.g.,][]{herwig99b, pignatari16, wagstaff20}.
However, \cite{lattanzio17} suggested that the strong buoyancy at the base of the convective pulse would lead to a small if not negligible amount of overshoot at this border, namely, $f_{\bot} \simeq 0$. 
In the recent work of \cite{karinkuzhi23}, the observation of two r/s-stars at metallicity [Fe/H]~$>-1$ was reported. Their abundance was suggested to originate from an AGB binary companion at [Fe/H]~$\simeq -0.5$ that experienced a PIE. This result was obtained by considering some overshoot at the top of the convective pulse in the AGB model ($f_{\top} = 0.06$) to facilitate the ingestion of protons and trigger i-process nucleosynthesis at higher mass and metallicities. 
The present paper details the extensive study we carried out on the impact of overshoot on the i-process in AGB stars. 

After introducing the necessary physics of the models in Sect.~\ref{sect:inputs}, we scrutinize the impact of a large set of overshoot parameters during the PIE of a 1~\Msun{} AGB model at [Fe/H]~$=-2.5$ (Sect.~\ref{sect:allparam}). In Sect.~\ref{sect:agbmz}, we examine the effect of a more limited set of overshoot parameters (four different $f_{\top}$ values), but in AGB models of various initial masses and metallicities. Section~\ref{sect:obs} is dedicated to the comparison of our models to the chemical abundances of observed stars. Our conclusions are given in Sect.~\ref{sect:concl}.

\begin{figure}[h!]
\centering
\includegraphics[width=1\columnwidth]{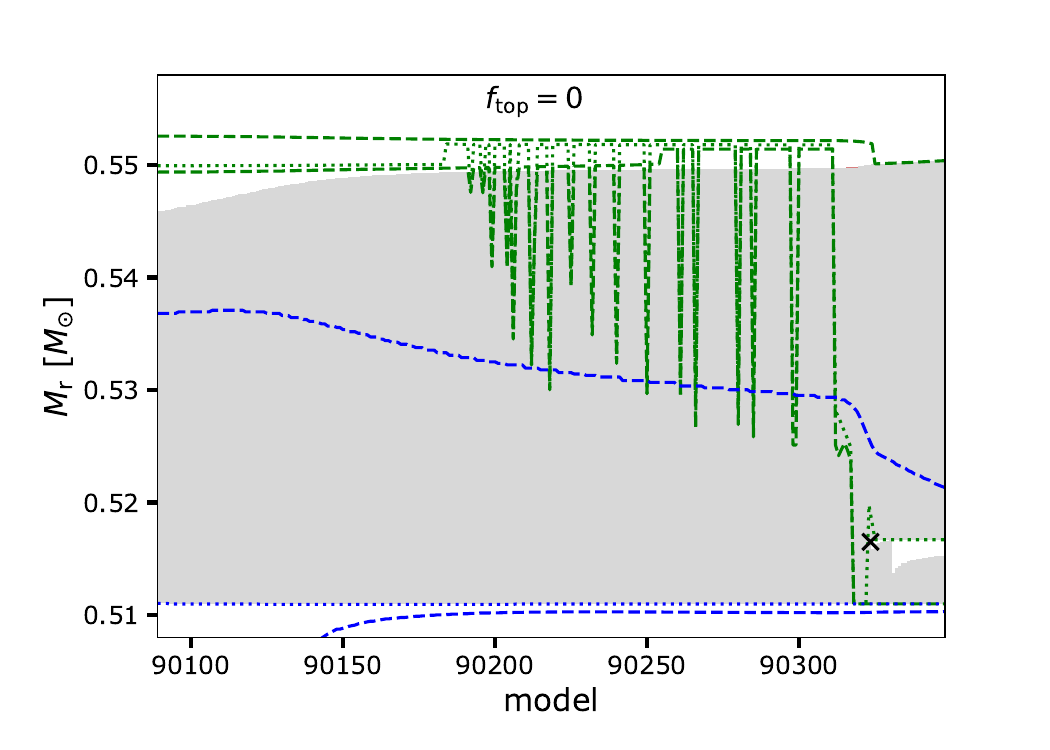}
\includegraphics[width=1\columnwidth]{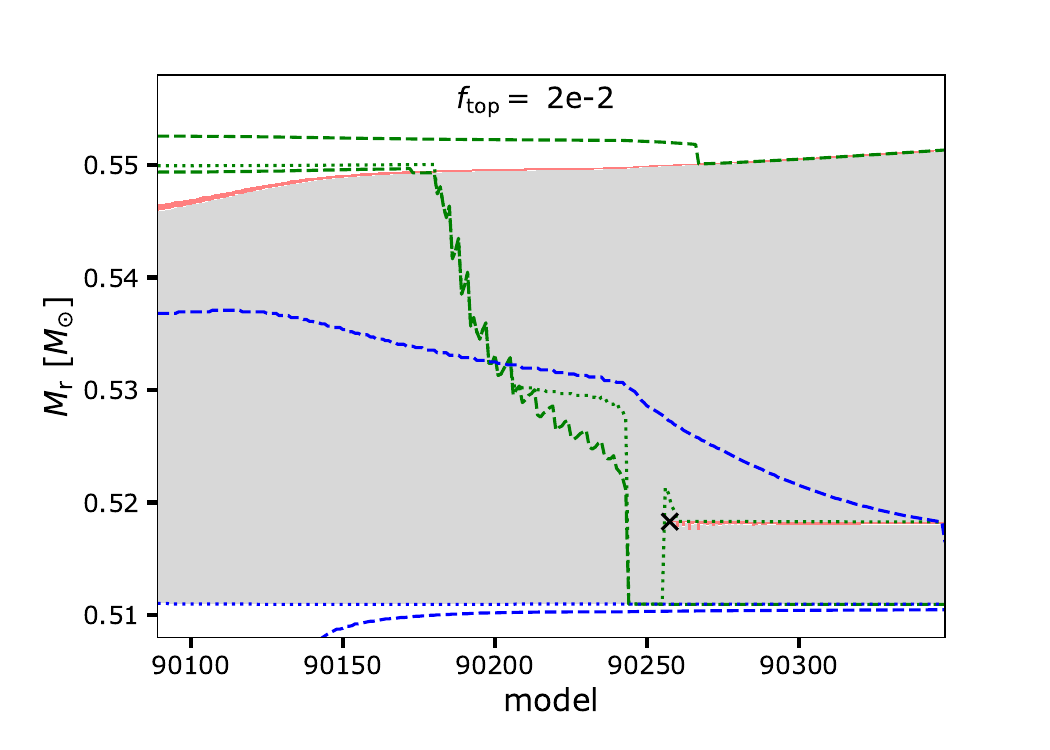}
\includegraphics[width=1\columnwidth]{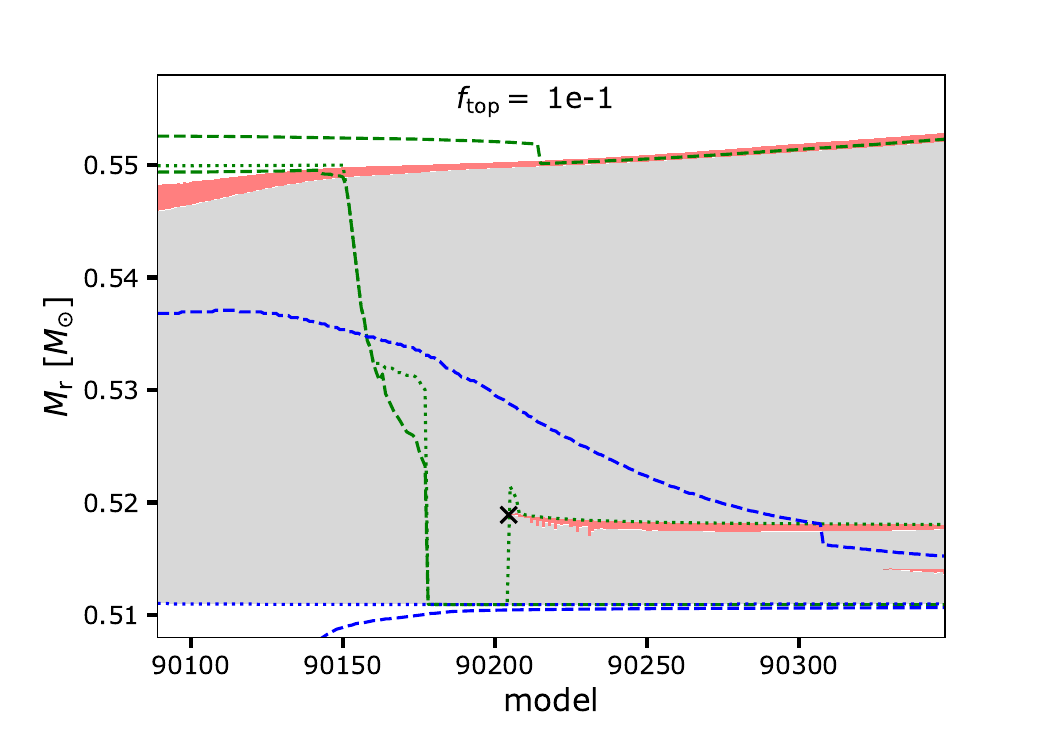}
\caption{Kippenhahn diagrams showing the PIE in a 1 \Msun, [Fe/H]~$=-2.5$ AGB model for three different values of the overshoot coefficient at the top of the convective pulse ($f_{\top}$).
Convective regions are shaded gray. 
The dotted green and blue lines trace the mass coordinate where the nuclear energy production by hydrogen and helium burning is maximum, respectively. 
The dashed green and blue lines delineate the hydrogen and helium-burning zones (where the nuclear energy production by H and He burning exceeds 10~erg~g$^{-1}$~s$^{-1}$). The red areas show the extent of the overshoot zones. The black crosses indicates where and when the convective pulse splits. 
}
\label{fig:kip_ftop}
\end{figure}

\begin{table}[t]
\scriptsize{
\caption{Overshoot coefficients used in the literature, at the bottom of the convective envelope ($f_{\env}$) at the top of the convective pulse ($f_{\top}$) and at the bottom of the convective pulse ($f_{\bot}$). 
\label{table:ff}
}
\begin{center}
\resizebox{8cm}{!} {
\begin{tabular}{lccc} 
\hline
Reference & $f_{\env}$ & $f_{\top}$  & $f_{\bot}$    \\
\hline
\cite{herwig97}   &   0.02  & 0.02  &  0.02    \\
\cite{herwig00}   &   0.016  & 0.016  &  0.016    \\
\cite{lugaro03}   &   0.128  & 0.016  &  0.016    \\
\cite{straniero06}   &   0.2  & $-$  &  $-$    \\
\cite{herwig07}   &   $-$  &  0.1 &   0.01 / 0.14   \\
\cite{weiss09}   &   0.016  & 0.016  &  0.016    \\
\cite{cristallo09b}   &   0.2  & $-$  &  $-$    \\
\cite{pignatari16}   &   0.126  & 0.014  &  0.008    \\
\cite{ritter18b}   &   0.126  & 0.014  &  0.008    \\
\hline
\end{tabular}
}
\end{center}
}
\end{table}


\section{Physical inputs of the models}
\label{sect:inputs}

The models were computed with the stellar evolution code STAREVOL \citep[][and references therein]{siess00, siess06, goriely18c}. 
We considered initial masses, at the zero-age main sequence (ZAMS), of $M_\mathrm{ini} =  1$, 2, 3, and 4 \Msun{} and metallicities in the range of $-2.5 \le $~[Fe/H]~$\le 0$ with the solar mixture of \cite{asplund09}.
We did not consider $\alpha-$enhanced mixtures for the sake of homogeneity. Discussions on the impact of an $\alpha-$enhancement on the PIE can be found in \cite{cristallo16} or \citet[][see their Sect. 4.7 in particular]{choplin22a} for instance.
The mass loss {prescription} of \cite{schroder07} is used up to the start of the AGB phase and then switched to the one of \cite{vassiliadis93}. 
As in the previous papers of this series, when the star becomes carbon rich, the opacity change due to the formation of molecules is taken into account following \cite{marigo02}. 
The mixing length parameter is set to 1.75. 
A network of 411 nuclei is used up to the occurrence of a PIE. When a PIE was about to start, we switched to an i-process network of 1160 nuclei, which we coupled to the transport equations.
We refer to \cite{choplin21,goriely21, choplin22a} for more details on the input physics. 
In this work, the models were computed from the ZAMS {and for some models (see table~\ref{table:3}) up to the end of AGB phase.}  
Nevertheless, as discussed in Sect.~\ref{sect:agbmz}, the missing part of the AGB phase of some models is not expected to significantly affect the resulting surface chemical abundances.
{Convective overshoot is included from the start of the AGB phase. Details on the implementation can be found in the next section. }

\subsection{Convective overshooting}

The modeling of overshooting is performed according to the prescription of \cite{goriely18c}, where the overshoot diffusion coefficient, $D_{\rm over}$, follows the expression
\begin{equation}
D_{\rm over} (z) = D_{\rm min} \, \times \, \left( \frac{D_{\rm cb}}{D_{\rm min}} \right)^{(1-z/z^{*})^{p}}
\label{eq:os18}
,\end{equation}
where $z^{*} = f_{\rm over} \, H_p \, \ln(D_{\rm cb}) / 2$ is the distance over which mixing occurs, $D_{\rm min}$ is the value of the diffusion coefficient at the boundary, $z=z^{*}$, and $p$ is a free parameter controlling the slope of the exponential decrease of $D_{\rm over}$ with $z$ \citep[see Fig.~1 of][for an illustration of the effect of the parameter, $p$]{goriely18c}. 
Below $D_{\rm min}$, we assume that $D_{\rm over} = 0$.
The $f_{\rm over}$ parameter corresponds to the radial extent over which the mixing takes place. 
We note that the original prescription of \cite{herwig97} is recovered if $D_{\rm min} = 1$~cm$^2$\,s$^{-1}$, $f_{\rm over} = 0.02,$ and $p=1$.
We refer to \citet[][especially Sect. 2]{goriely18c} for more details.
In the present study, when not mentioned otherwise, we adopt the default values of $D_{\rm min}=1$~cm$^2$\,s$^{-1}$ and $p=1$.

\begin{figure}[t!]
\centering
\includegraphics[width=\columnwidth]{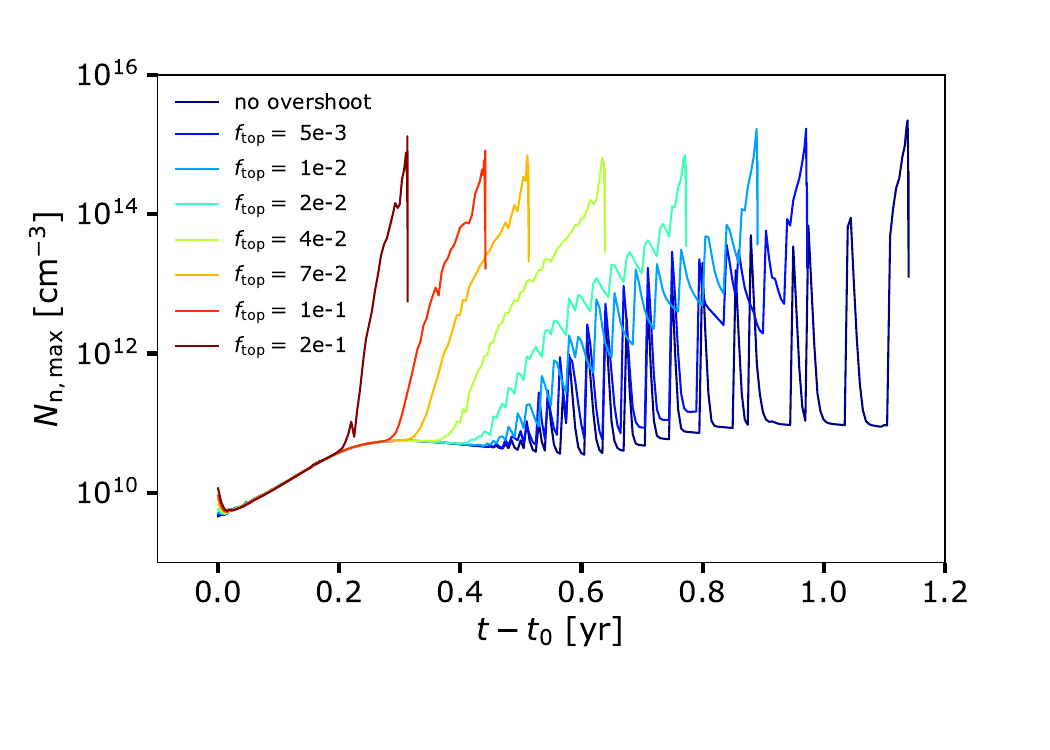}
\vspace{-1cm}
\caption{Evolution of maximal neutron density during a PIE. The different lines correspond to different values of the overshoot coefficient at the top of the convective pulse ($f_{\top}$).
{The evolution is illustrated between the time $t_0$ at which $N_{\rm n, max}$ rises above $10^{10}$~cm$^{-3}$ and ends when the convective pulse splits.}
}
\label{fig:nnmax_ftop}
\end{figure}

\begin{figure}[h!]
\centering
\includegraphics[width=\columnwidth]{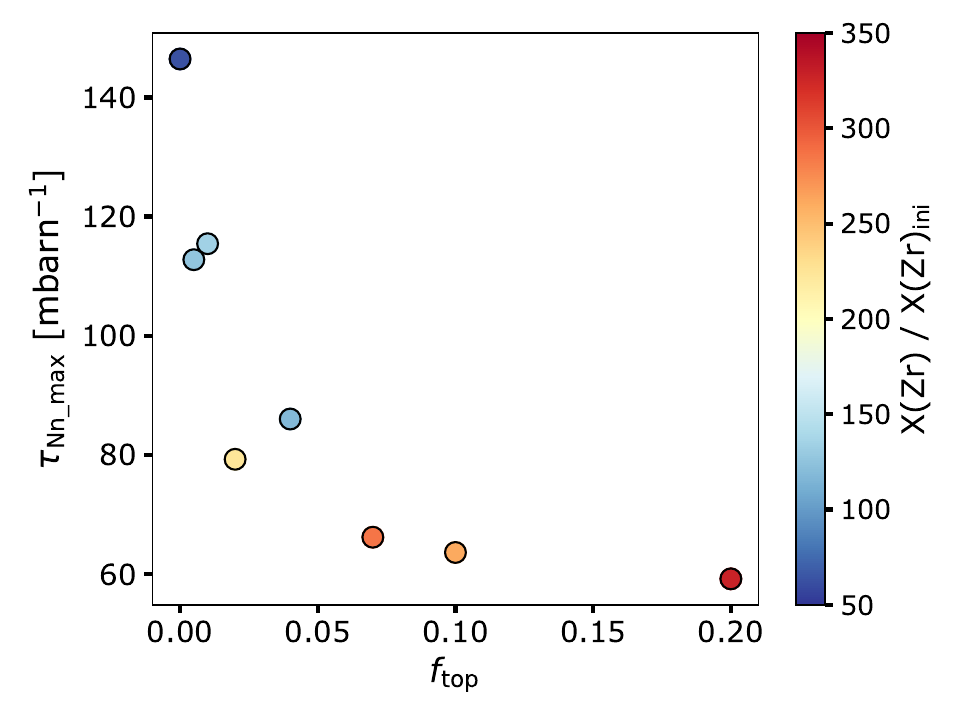}
\caption{Neutron exposure at the bottom of the convective pulse (where the neutron density, $N_n$, is maximum) as a function of the $f_{\top}$ parameter. The color indicates {the Zr ($Z=40$) production factor, namely, its surface abundance normalized to its initial abundance}.
}
\label{fig:taunnmax}
\end{figure}

\begin{figure*}[h!]
\centering
\includegraphics[width=\columnwidth]{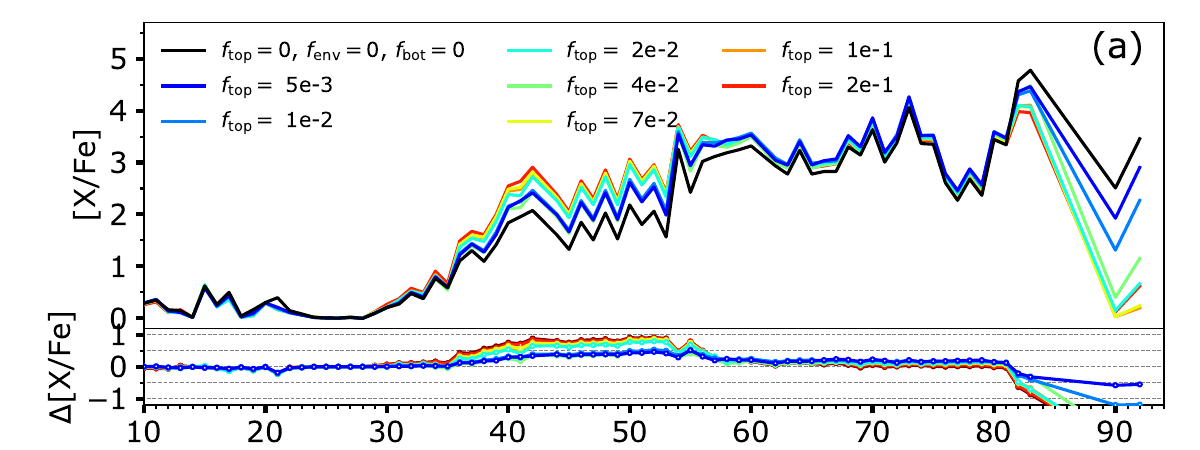}
\includegraphics[width=\columnwidth]{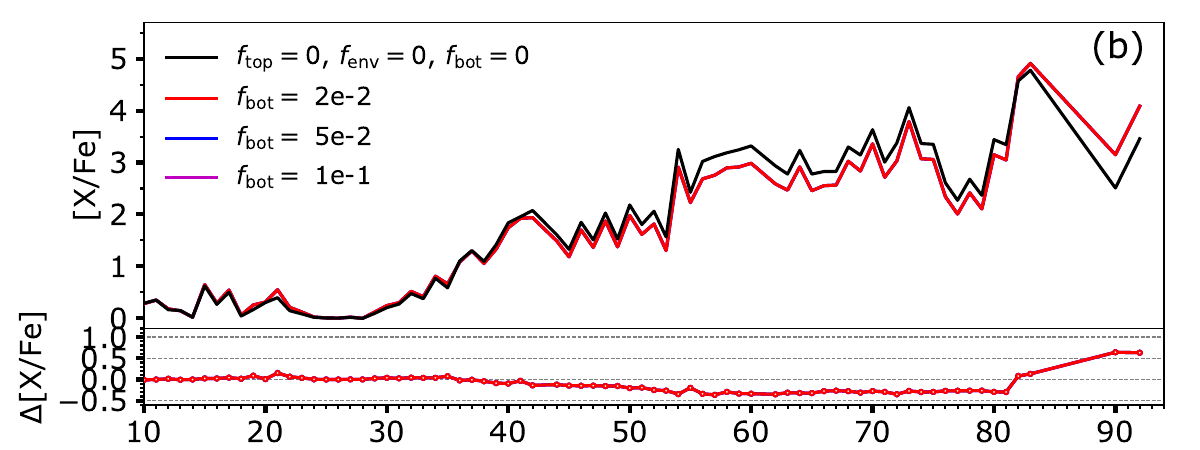}
\includegraphics[width=\columnwidth]{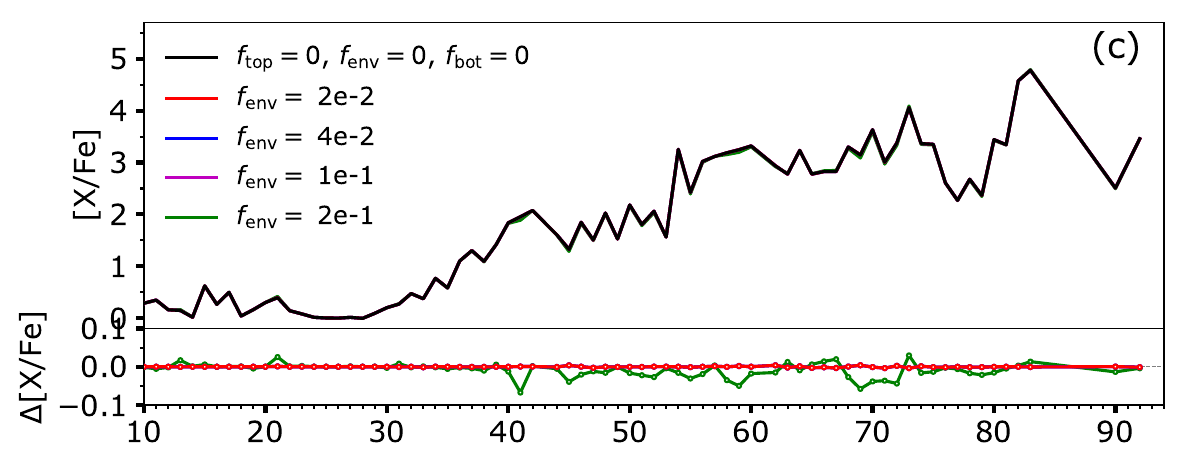}
\includegraphics[width=\columnwidth]{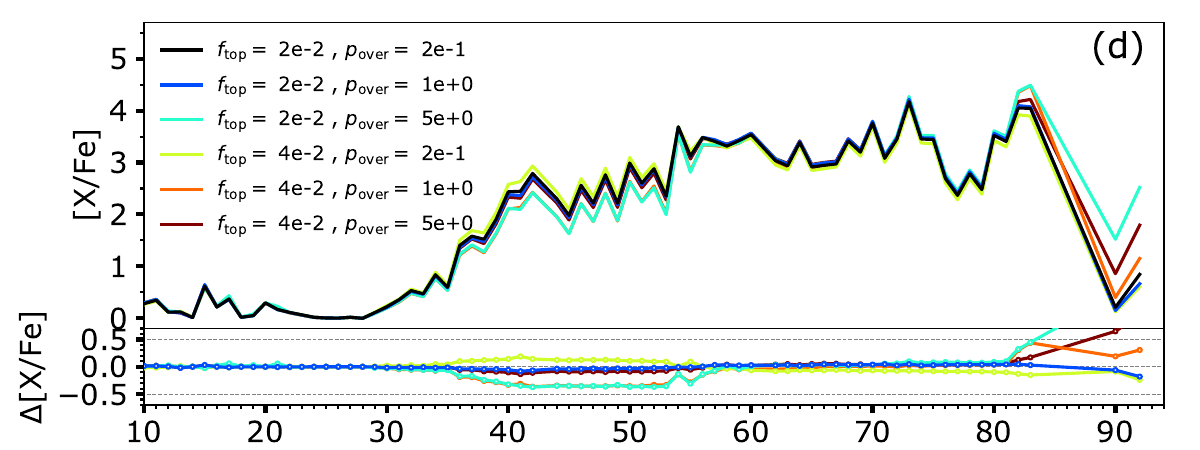}
\includegraphics[width=\columnwidth]{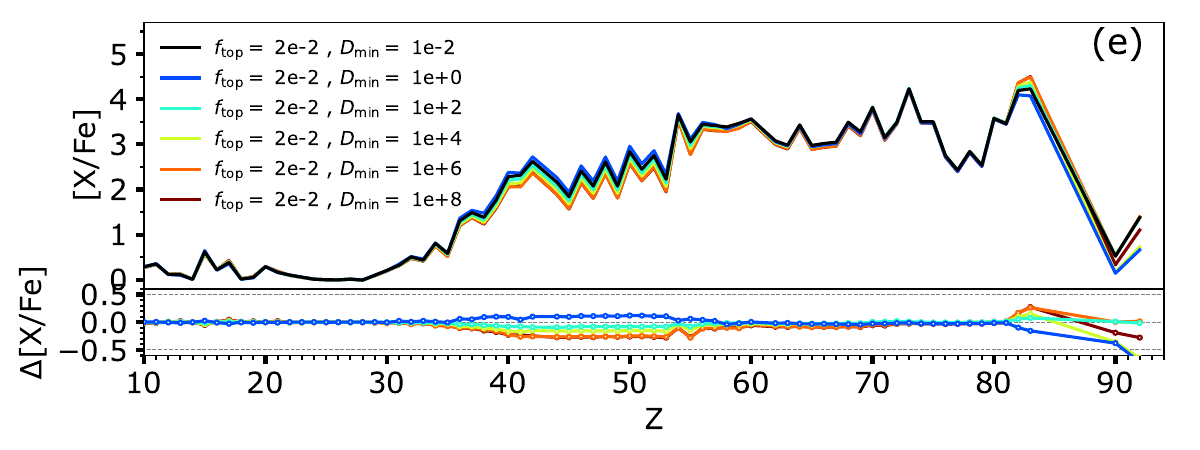}
\includegraphics[width=\columnwidth]{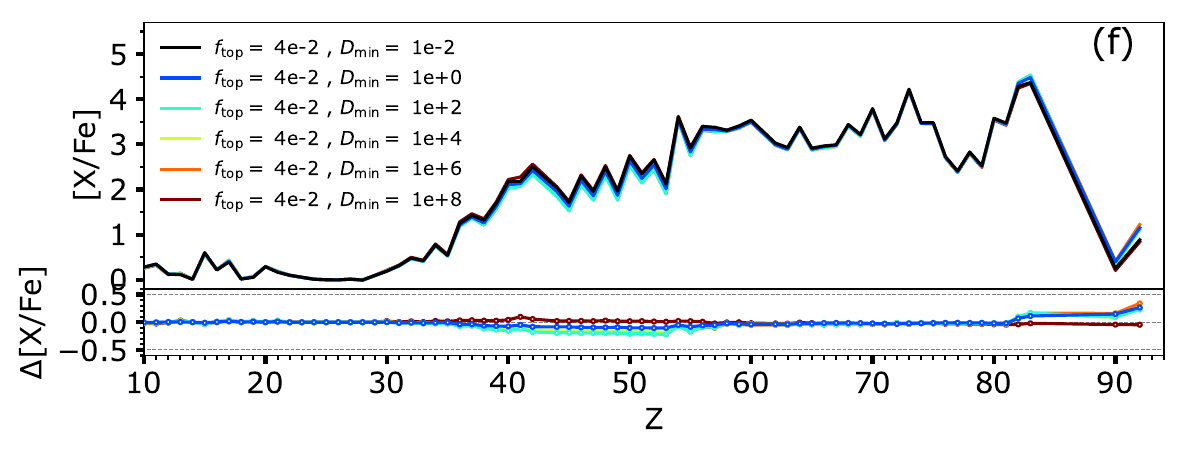}
\caption{Impact of different overshoot parameters on the surface [X/Fe] ratios after a PIE in a 1~\Msun, [Fe/H]~$=-2.5$ AGB model. Shown are the effect of including overshooting at the top of the convective pulse (panel a), at the bottom of the convective pulse (panel b), at the bottom of the convective envelope (panel c), when varying the parameters $p$ (with $f_{\top} = 0.02$ and $f_{\top} = 0.04$, panel d), and $D_{\rm min}$ with $f_{\top}=0.02$ (panel e), and $f_{\top}=0.04$ (panel f). 
The bottom subplots report the difference between a given model and the first model in the list.
}
\label{fig:xfe_os}
\end{figure*}

\begin{table}[h!]
\scriptsize{
\caption{Initial mass ($M_{\rm ini}$), [Fe/H] ratio  and metallicity $Z$ of our grid models.
\label{table:2}
}
\begin{center}
\resizebox{6cm}{!} {
\begin{tabular}{lcrc} 
\hline
\hline
Model & $M_{\rm ini}$  & [Fe/H] &$Z$      \\
 label  &    [\Msun]        &        &  \\
\hline
M1.0z2.0   &   1.0  & $-2.0$  &  1.4e-4  \\
M1.0z1.5   &  1.0   & $-1.5$  &  4.3e-4    \\
M1.0z1.4   &  1.0   & $-1.4$  &  5.4e-4    \\
M1.0z1.3   &  1.0   & $-1.3$  &  6.8e-4   \\
M1.0z1.2   &  1.0   & $-1.2$  &  8.6e-4   \\
M1.0z1.1   &  1.0   & $-1.1$  &  1.1e-3    \\
M1.0z1.0   &  1.0   & $-1.0$  &  1.4e-3   \\
M1.0z0.5   &  1.0   & $-0.5$  &  4.3e-3   \\
M1.0z0.0   &  1.0   & $0.0$   &  1.4e-2   \\
\hline
M1.5z1.3   &  1.5   & $-1.3$   &  6.8e-4   \\
M1.5z1.2   &  1.5   & $-1.2$   &  8.6e-4   \\
\hline
M1.7z1.4   &  1.7   & $-1.4$   &  5.4e-4   \\
\hline
M1.8z1.5   &  1.8   & $-1.5$   &  4.3e-4   \\
\hline
M2.0z2.0   &  2.0   & $-2.0$  &  1.4e-4   \\
M2.0z1.5   &  2.0   & $-1.5$  &  4.3e-4   \\
M2.0z1.0   &  2.0   & $-1.0$  &  1.4e-3   \\
M2.0z0.5   &  2.0   & $-0.5$  &  4.3e-3    \\
M2.0z0.0   &  2.0   & $0.0$   &  1.4e-2   \\
\hline
M3.0z2.5   &  3.0   & $-2.5$  &  4.3e-5   \\
M3.0z2.0   &  3.0   & $-2.0$  &  1.4e-4   \\
\hline
M4.0z2.0   &  4.0   & $-2.0$  &  1.4e-4    \\
\hline 
\end{tabular}
}
\end{center}
}
\end{table}

\begin{table*}[h!]
\scriptsize{
\caption{Properties of the computed models for the various values of $f_\top$. In the column PIE, we specify if a PIE occurs. $N_\mathrm{p}$ gives the total number of computed thermal pulses, $\#$ the pulse number at which the PIE occurs, $N_{\rm n, max}$  (in cm$^{-3}$) the maximal neutron density during that event {and $M^{\rm env}_{\rm fin}$ the convective envelope mass left at the end of the calculation (in $M_{\odot}$). A '$+$' sign in the $N_\mathrm{p}$ column indicates that additional pulses may develop since the simulation was stopped before the end of the AGB phase. }
\label{table:3}
}
\begin{center}
\resizebox{18.5cm}{!} {
\begin{tabular}{l|ccc|clcc|clcc|clcc} 
\hline
\hline
Model & \multicolumn{3}{|c|}{$f_{\top}=0$}   & \multicolumn{4}{|c|}{$f_{\top}=0.02$}  & \multicolumn{4}{|c|}{$f_{\top}=0.04$}   &   \multicolumn{4}{|c}{$f_{\top}=0.10$}     \\
 label  & PIE & $N_{\rm p}$   & $M^{\rm env}_{\rm fin}$ &   PIE  &   \#\,/\,$N_{\rm p}$  & $N_{\rm n, max}$  & $M^{\rm env}_{\rm fin}$ & PIE  &   \#\,/\,$N_{\rm p}$ & $N_{\rm n, max}$ & $M^{\rm env}_{\rm fin}$ & PIE  &   \#\,/\,$N_{\rm p}$       & $N_{\rm n, max}$ & $M^{\rm env}_{\rm fin}$                   \\
\hline
M1.0z2.0   &NO& 4   & {0.01} & YES &   2\,/\,2   &7.2e14& {0.02} &YES&   {1\,/\,1} &1.6e14 &{0.03} & YES  & 1\,/\,1 & 2.1e14 & {0.01}   \\
M1.0z1.5   &NO& 3   & {0.01} & NO   & $-$  &  $-$    &$-$ &YES& {2\,/\,2}  &5.6e14 & {0.01} & YES  &  1\,/\,1  & 1.8e14& {0.02}  \\
M1.0z1.4   &NO& 4   & {0.01} & NO   & $-$  &  $-$    &$-$ &NO  &$-$& $-$    &$-$ & YES  &  1\,/\,1  & 2.1e14  & {0.02} \\
M1.0z1.3   &NO& 4   & {0.01} & NO   & $-$  &  $-$    &$-$ &NO  &$-$& $-$    &$-$ & YES  &  1\,/\,1  & 2.7e14 & {0.03} \\
M1.0z1.2   &NO& 3   & {0.01} & NO   & $-$  &  $-$    &$-$ &NO &$-$& $-$    &$-$ & YES  &  1\,/\,1  & 5.4e14 & {0.03} \\
M1.0z1.1   &NO& 4   & {0.01} & NO   & $-$  &  $-$    &$-$ &NO  &$-$& $-$   &$-$  & YES  &  1\,/\,1  & 1.4e14 & {0.02} \\
M1.0z1.0   &NO& 3   & {0.01} & NO   & $-$  &  $-$    &$-$ &NO  &$-$& $-$   &$-$  & YES  & 1\,/\,1   &   1.3e14 & {0.02} \\
M1.0z0.5   &NO& 5   & {0.02} & NO   & $-$  &  $-$    &$-$ &NO  &$-$& $-$   &$-$  & YES  &  1\,/\,1  & 5.4e13 & {0.04}\\
M1.0z0.0   &NO& 5   & {0.02} & NO   & $-$  &  $-$    &$-$ &NO  &$-$& $-$   &$-$  & NO    & $-$    & $-$   &$-$  \\
\hline
M1.5z1.3$^*$&NO& {8+}   & {0.18} &   &      &        &               &     &     &        &   & YES   & 1\,/\,1+& 1.8e14  & {0.05}  \\
M1.5z1.2$^*$   &NO& 6   & {0.04} &   &      &        &               &     &     &       &    & YES   & 1\,/\,3& 1.6e14 & {0.01}   \\
\hline
M1.7z1.4$^*$   &NO& 7+  &{0.63} &  &      &        &              &     &     &       &    & YES   & 1\,/\,4& 4.0e14  & {0.01}  \\
\hline
M1.8z1.5$^*$   &NO& 6+& {0.85} &    &      &         &              &     &     &          &  & YES   & {1\,/\,6}& 2.4e14  & {0.02} \\
\hline
M2.0z2.0   &NO& 13 & {0.05} &  NO  & $-$    & $-$    &$-$ &YES& 2\,/\,3+  & 1.3e15 & {1.01} & YES & {1\,/\,7+}  &  3.8e14 & {0.65} \\
M2.0z1.5   &NO& 12 & {0.02} &  NO  & $-$    & $-$    &$-$ &NO  & $-$    & $-$    &$-$ & YES &  1\,/\,6  &  4.7e14 & {0.02} \\
M2.0z1.0   &NO& 11 & {0.09} &  NO  & $-$    & $-$    &$-$ &NO   & $-$    & $-$    &$-$ & YES &  1\,/\,9  &  1.1e14& {0.06}  \\
M2.0z0.5   &NO& 11 & {0.04} &  NO  & $-$    & $-$    &$-$ &NO   & $-$    & $-$    &$-$ & YES &  2\,/\,6  & 8.2e13 & {0.05} \\
M2.0z0.0   &NO& 12 & {0.02} &  NO  & $-$    & $-$    &$-$ &NO   & $-$    & $-$    &$-$ & NO &  $-$  &  $-$ &  $-$ \\
\hline
M3.0z2.5   &NO& {20+} & {0.24} &  NO & $-$    & $-$    &$-$ &YES& 3\,/\,5+  & 1.2e15 & {1.70} & YES &  2\,/\,6+  &  2.0e15 & {1.65} \\
M3.0z2.0   &NO& {21+} & {0.12} &  NO & $-$    & $-$    &$-$ &NO   & $-$    & $-$    &$-$ & YES & 2\,/\,7+  &   8.1e14 & {1.67} \\
\hline
M4.0z2.0   &NO& 28+ & {0.34} &  NO & $-$    & $-$    &$-$ &NO   & $-$    & $-$    &$-$ &  NO & $-$    &  $-$    &$-$   \\
\hline 
\end{tabular}
}
\end{center}
}
\normalsize{
$^*$ The possible occurrence of PIEs in these models for $f_{\top} = 0.02$ and 0.04 was not investigated.
}
\end{table*}


\section{Impact of overshoot in a 1~\Msun{} [Fe/H]~$=-2.5$ AGB model }
\label{sect:allparam}

In this section, we discuss the impact of overshoot during the PIE of a 1~\Msun, [Fe/H]~$=-2.5$ AGB model. 
We consider overshooting above ($f_{\top}$) and below ($f_{\bot}$) the convective pulse, as well as below the convective envelope ($f_{\env}$). 
We also look at the impact of the $D_{\rm min}$ and $p$ parameters (cf. Eq.~\ref{eq:os18}).
In total, 30 AGB models during a given PIE were computed. These calculations being computationally expensive, we use the same dilution procedure as in \citet[][cf. Sect. 3.1]{martinet23} to determine the surface composition after the PIE, which is briefly recalled below.

\subsection{The dilution procedure}

In this procedure, models are stopped when the convective pulse splits, that is, just after the neutron density peak. The final surface abundances are predicted by diluting the chemical abundances in the homogenized pulse with those of the envelope. As discussed in \cite{martinet23}, this method leads to a very accurate estimate of the final surface abundances with the exception of Li, C, and N. In this work, we once again checked  the accuracy of this procedure by computing several models up to the merging of the pulse with the envelope. 
When comparing the final abundances to the ones derived from the dilution procedure, a maximal deviation of about 10~\% is noted.
This procedure is only used in the present Sect.~\ref{sect:allparam} to analyze the impact of the various overshoot parameters.

\subsection{The impact of overshooting above the convective thermal pulse}

We first included the overshoot at the top of all convective boundaries. Eight values of $f_{\top}$ were considered, from 0 to 0.2, with $D_{\rm min} = 1$~cm$^2$~s$^{-1}$ and $p=1$. 
Without overshoot or with small values of $f_{\top}$, protons are not ingested continuously because of the strong chemical discontinuity existing between the convective pulse and the intershell radiative zone. As the top of the convective pulse advances in mass, protons are engulfed but at the start of the PIE they are burnt before reaching the deepest layers. 
This erosion of the base of the H-burning shell leads to successive minor PIEs, producing the spikes seen in the location of the nuclear energy production by H-burning (green lines in the top panel of Fig.~\ref{fig:kip_ftop}) and in the maximal neutron density (Fig.~\ref{fig:nnmax_ftop}). 
{This spiky behavior persists with increasing spatial and temporal resolution. To our understanding, this is an almost unavoidable feature resulting from the discretization of the models and in particular, to the discontinuity at the convective boundary in the absence of extra mixing. Without additional transport processes, the upper boundary of the convective pulse grows discretely and erodes the base of the H-rich shell (and mix protons in the pulse) intermittently. 
These spikes might nevertheless disappear for very high spatial and temporal resolutions, as the pulse grows more progressively and the erosion becomes more gradual. 
}
When the pulse has sufficiently grown, the amount of proton engulfed becomes high enough to trigger the proper PIE with neutron densities of $N_{\rm n} \sim 10^{15}$~cm$^{-3}$. The pulse then splits and eventually  merges with the envelope. As seen in Fig.~\ref{fig:kip_ftop} (top panel), the proper PIE starts around model 90320. 
Nevertheless, we note that the previous minor ingestions of protons, reaching $10^{12}$~cm$^{-3} < N_{\rm n} < 10^{14}$~cm$^{-3}$ (Fig.~\ref{fig:nnmax_ftop}), already lead to the production of trans-iron elements.

Increasing $f_{\top}$ reduces the chemical discontinuity between the convective pulse and the intershell radiative zone. As a consequence, the spikes seen in the nuclear energy production (Fig.~\ref{fig:kip_ftop}) are removed and the profile of $N_{\rm n, max}$ (Fig.~\ref{fig:nnmax_ftop}) becomes smoother. 
{The PIE also starts earlier because the growing convective pulse reaches the H-rich layers earlier (this is visible in Fig.~\ref{fig:nnmax_ftop}, where the $N_{\rm n}$ profiles starts to rise earlier with increasing $f_{\rm top}$). 
In addition, the duration of the PIE\footnote{{The duration of the PIE is arbitrarily defined from the time the maximal neutron density first rises above $10^{10}$~cm$^{-3}$ until the convective pulse splits. 
}
} is shortened with increasing $f_{\rm top}$. Indeed, for high $f_{\rm top}$ values, the PIE starts when the top of the pulse grows fast in mass; in this case, the H-rich layers that will bring the large amount of protons responsible for the pulse to split are quickly reached. In contrast, for lower $f_{\rm top}$ values, the PIE starts later, when the pulse grows more slowly, so that it takes more time to engulf the critical amount of H needed for the splitting the convective pulse. }

{Shorter PIEs translate into smaller neutron exposures\footnote{The neutron exposure $\tau$ is computed as 
$\tau \, = \int  N_{\rm n} \, v_{\rm T} \, dt$ where $N_{\rm n}$ is the neutron density and $v_{\rm T} = \sqrt{ \, 2 \, k_B \, T /m_{\rm n}}$ the neutron thermal velocity with $k_B$ the Boltzmann constant, $T$ the temperature fixed to 250~MK, and $m_n$ the neutron mass.}  and a higher production of elements in the Zr region, as illustrated in Fig.~\ref{fig:taunnmax}. 
} 
With smaller neutron exposures the production of $36 \leq Z \leq 56$ elements increases to the detriment of elements with $Z \ge 82$ and, in particular, of the actinides (\ref{fig:xfe_os}a).
{As discussed in \citet{choplin22b}, a neutron density of at least $\sim 1.5 \times 10^{15}$~cm$^{-3}$ is required to synthesize actinides}. 
In the 1~\Msun{} model, this threshold value is obtained for low $f_{\top}\sim 0.02$.

\begin{figure*}[t]
\centering
\includegraphics[width=2\columnwidth]{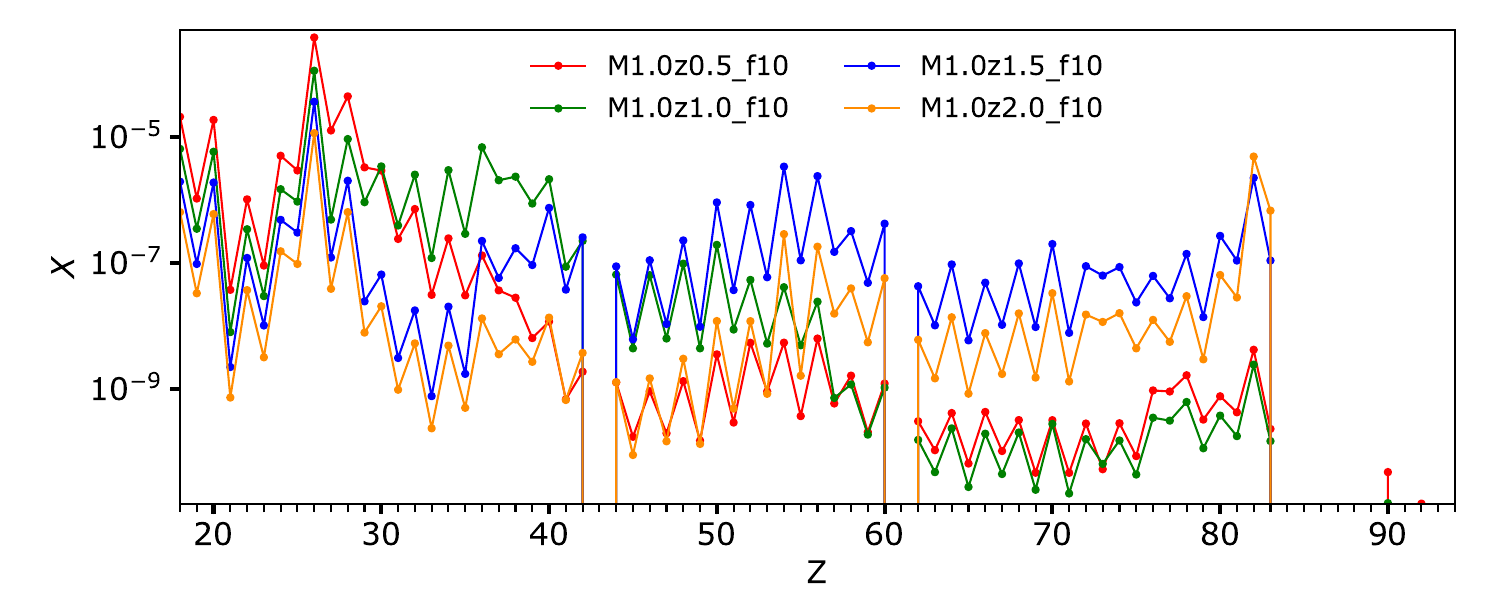}
\caption{Final surface mass fractions (after decays) for the 1 $M_{\odot}$ that experiences a PIE.  The different colors correspond to the metallicities $-0.5$ (red), $-1$ (green), $-1.5$ (blue) and $-2$ (orange). Models were computed with $f_{\top} = 0.1$. 
}
\label{fig:abund_mf}
\end{figure*}

\subsection{The impact of overshooting at other convective boundaries}

As a next step, we considered overshoot only at the bottom of the convective pulse, varying $f_{\bot}$ from 0.02 to 0.1. As seen in Fig.~\ref{fig:xfe_os}(b), setting $f_{\bot} \neq 0$ produces small changes in the surface abundances by at most 0.5~dex. The elemental distribution is almost unaffected by changes in  $f_{\bot}$ as long as it is non-zero.
Finally, we include overshoot only at the bottom of the convective envelope and varied $f_{\env}$ between 0.02 and 0.2. Here, again the impact on the surface abundances is very small ($<0.1$ dex, Fig.~\ref{fig:xfe_os}c).
In the end, the abundances are mostly sensitive to $f_{\top}$ since this  parameter directly controls how protons are engulfed in the convective pulse.

\subsection{The impact of the $p$ and $D_{\rm min}$ parameters}

Finally, we investigate how the other two free parameters, $p$ and $D_{\rm min}$, which control the profile of the diffusion coefficient (Eq.~\ref{eq:os18}), impact surface abundances. 
In a first step, $p$ was set to 0.2, 1 or 5, with $f_{\top} = 0.02$ or 0.04 and $D_{\rm min}=1$ (6 models). 
In a second step, $D_{\rm min}$ was varied from $10^{-2}$ to $10^{8}$~cm$^2$\,s$^{-1}$, with $f_{\top} = 0.02$ or 0.04 and $p=1$ (12 models).
In both sets of models, the abundance scatter is on the order of 0.5~dex at maximum for elements with $35 < Z < 55$ and for Pb and Bi (Fig.~\ref{fig:xfe_os}, panels d, e, and f). Furthermore, Th and U show greater variations and a strong dependence with $p$ ($\sim 1.5 - 2$ dex at maximum).

\subsection{Summary}

Varying the overshoot parameters in our 1\Msun{} [Fe/H]=-2.5 model produces variations in the abundances of $36<Z<56$ elements of $0.5-1$ dex. The scatter is $2-3$ dex for Th and U. The other elements are affected by less than 0.5 dex. 
Nuclei with $56<Z<80$ have a higher predictive power because they are weakly affected by the changes in the overshoot parameters (Fig.~\ref{fig:xfe_os}).
Although the nucleosynthesis may be affected, variations in the overshoot parameters have a weak impact of the structure of the PIE (provided it is present without overshooting).

\begin{figure}[t]
\centering
\includegraphics[width=0.98\columnwidth]{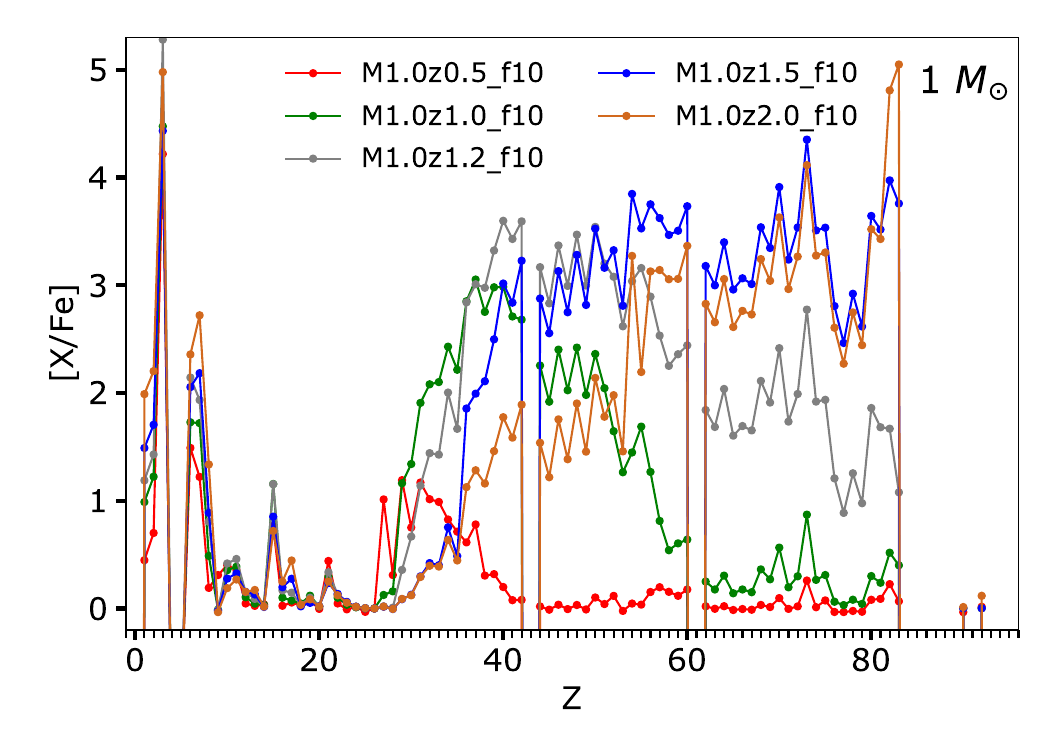}
\includegraphics[width=0.98\columnwidth]{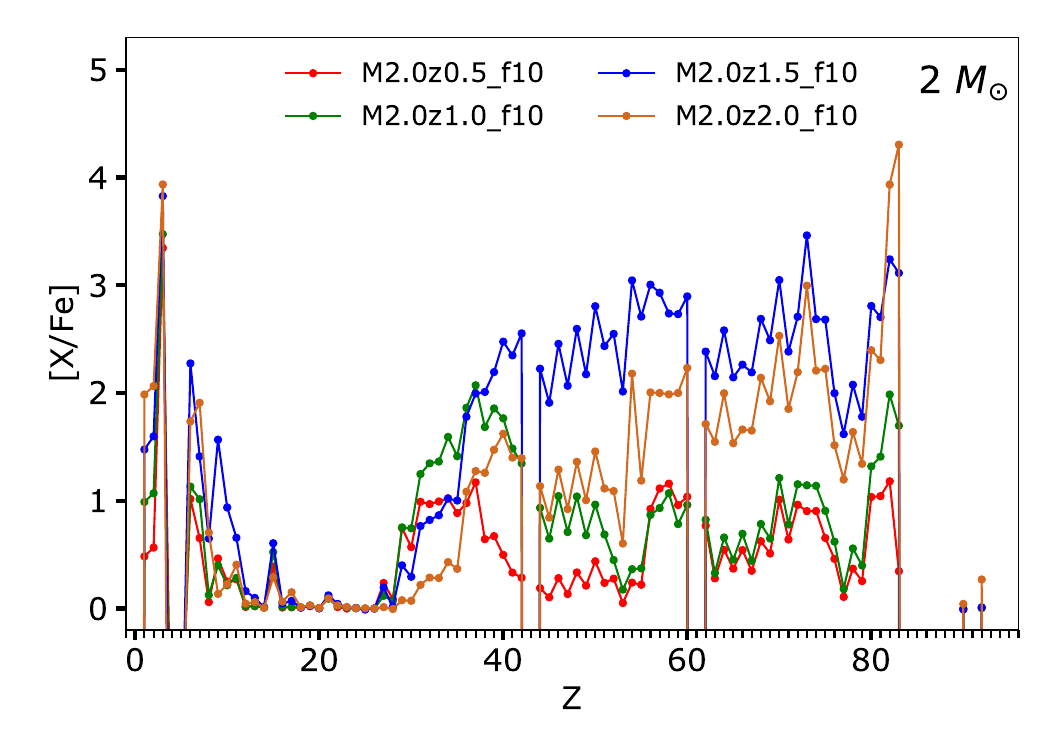}
\includegraphics[width=0.98\columnwidth]{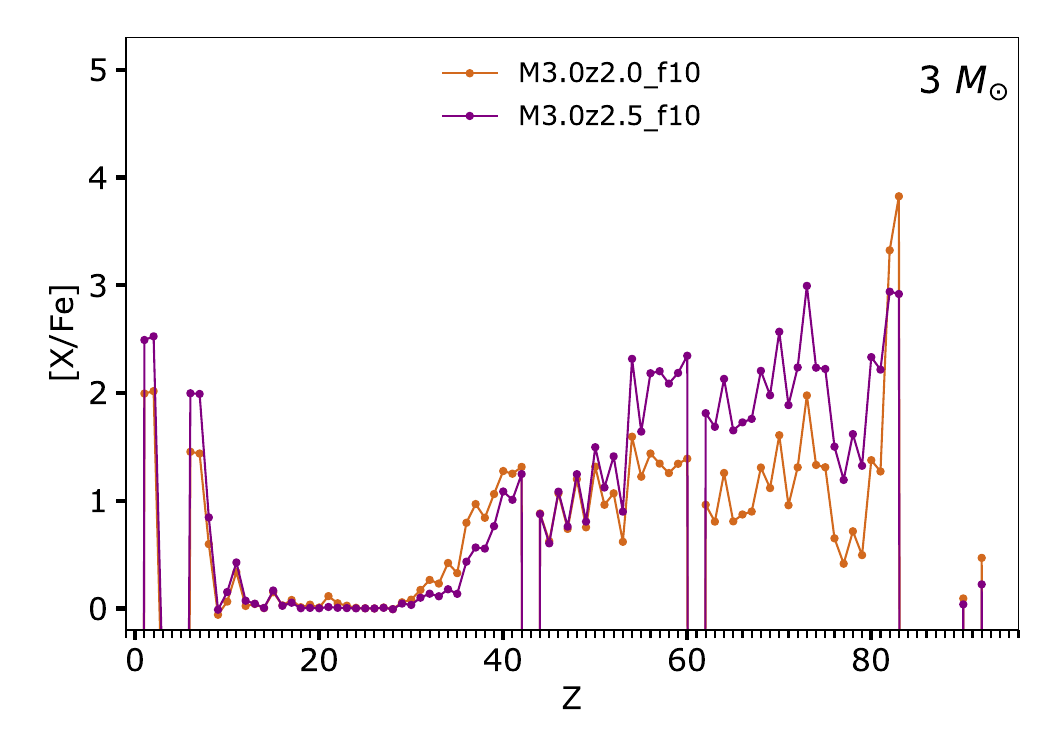}
\caption{Final surface [X/Fe] ratios for the models that experience a PIE. The different panels correspond to different initial stellar  masses. The various colors correspond to different metallicities [Fe/H]~$-0.5$ (red), $-1$ (green), $-1.2$ (grey), $-1.5$ (blue), $-2$ (dark orange), and $-2.5$ (purple). All these models were computed with $f_{\top} = 0.1$ A few models are not shown for clarity. 
}
\label{fig:abund_xfe}
\end{figure}


\section{AGB models at different masses and metallicities with overshoot}
\label{sect:agbmz}

We computed models of 1, 2, 3, and 4~\Msun{} with metallicities in the range of $-2<$~[Fe/H]~$\le 0$ (Table~\ref{table:2}).
For each model, we considered four overshoot cases: no overshoot and overshoot above the pulse with $f_{\top} = 0.02$, $f_{\top} = 0.04,$ and $f_{\top} = 0.1$. These values scan the range of $f_{\top}$ used thus far in the literature ($0.014<f_{\top}<0.1$, cf. Table~\ref{table:ff}). We set $f_{\env} = 0$, $f_{\bot}= 0$, $D_{\rm min} = 1$~cm$^{-3}$ and $p=1$.
The models are labeled as MX.XzY.Y\_fZZ, where X.X corresponds to the mass in~\Msun, Y.Y~$= -$~[Fe/H] and ZZ is the value of $f_{\top}$. For instance, M1.0z2.0\_f04 refers to the 1~\Msun{} model at [Fe/H]~$=-2.0$ computed with $f_{\top} = 0.04$.
{When possible, the models were computed until the end of the TP-AGB phase. Some models were stopped before the end (cf. Table~\ref{table:3} for more details).}

\subsection{Evolution and nucleosynthesis of the 1~\Msun{} models}
\label{sect:nuc1msun}

Overshoot at the top of the convective pulses facilitates the development of a PIE. The higher $f_{\top}$, the easier for a PIE to develop. Setting $f_{\top} = 0.02$ triggers a PIE in the 1~\Msun{} [Fe/H]~$=-2$ model only (Table~\ref{table:3}). With $f_{\top} = 0.04$, a PIE develops in the [Fe/H]~$=-2$ and $-1.5$ models. Finally, $f_{\top} = 0.1$ gives a PIE in all 1~\Msun{} models, except at solar metallicity.

In all the  1~\Msun{} model where a PIE develops, a large amount of carbon (and heavy elements) is dredged-up to the surface, considerably increasing the  opacity of the envelope. This boosts the mass-loss rate leading to the ejection of  the entire envelope before another thermal pulse appears \citep[cf. Sect.~3.1.4 in][for more details]{choplin21}. In all our 1~\Msun{} models, the AGB phase quickly ends after a PIE.

Figure~\ref{fig:abund_mf} shows the final surface mass fractions of the 1~\Msun{} models with $f_{\top} = 0.1$ that experience a PIE.
Generally speaking, the lower the metallicity, the heavier the elements synthesized. 
The main reason is that the abundance of $^{56}$Fe which acts as the main seed, increases with metallicity, so that the neutron to seed ratio decreases with increasing metallicity. 
More $^{56}$Fe favors the production of lighter i-process elements to the detriment of the heavier ones. A similar metallicity dependence on the distribution of heavy elements is found for the standard s-process nucleosynthesis \citep{gallino98,goriely00,goriely18c}.
More specifically, at a metallicity of [Fe/H]~$=-0.5$ ($-1.0$), only elements with $26 \lesssim Z \lesssim 40$ ($26 \lesssim Z \lesssim 55$)  are synthesized (Fig.~\ref{fig:abund_mf}). 
In contrast, at [Fe/H]~$=-1.5$ (blue pattern, in Fig.~\ref{fig:abund_mf}), nuclei with $Z \gtrsim 55$ start to be  significantly produced. 
Going down to [Fe/H]~$=-2.0$ (orange pattern) leads to a large production of heavy elements (especially Pb and Bi) to the detriment of lighter elements. 
No significant Th and U are synthesized in these models because of the high $f_{\top}$ value (cf. Sect.~\ref{sect:allparam}). 
The final surface [X/Fe] ratios are displayed in the top panel of Fig.~\ref{fig:abund_xfe}. 
The [Fe/H]~$=-1.5$ and $-2.0$ models show a strong i-process signature with large overproduction factors up to $4-5$ dex. 
As discussed previously, increasing the metallicity leads to smaller heavy elements [X/Fe] ratios.

\begin{figure*}[t]
\centering
\includegraphics[width=2.1\columnwidth]{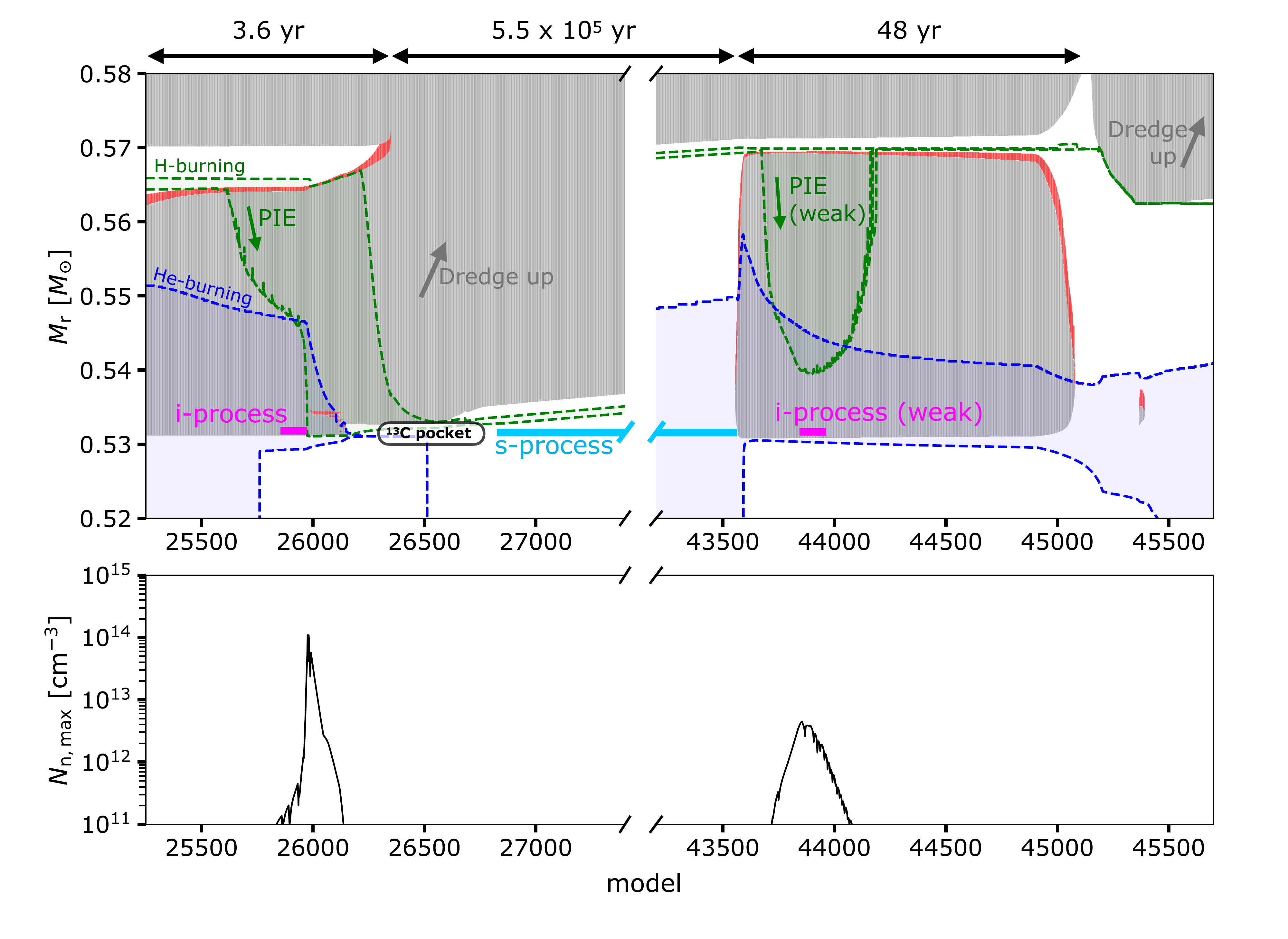}
\vspace{-0.5cm}
\caption{Kippenhahn diagram showing the early AGB phase (first and second thermal pulses) of a 2 \Msun, [Fe/H]~$=-1.0$ star (M2.0z1.0\_f10 model). 
Convective regions are shaded gray. 
The dashed green and blue lines delineate the hydrogen and helium-burning zones, respectively (where the nuclear energy production by H and He burning exceeds 10~erg~g$^{-1}$~s$^{-1}$). 
The red area shows the extent of overshooting. 
The magenta (cyan) line indicates the region zones where the i-process (s-process) nucleosynthesis occurs.
The bottom panel shows the maximal neutron density as a function of the model number, or equivalently time as indicated in the upper x-axis.
}
\label{fig:kip_m1.0z2.0}
\end{figure*}

\subsection{Evolution and nucleosynthesis of the 2 \Msun{} models}

The 2~\Msun{} models experiencing a PIE show a more complex evolution and nucleosynthesis than the 1~\Msun{} models: after a PIE, the AGB phase resumes with the occurrence of additional thermal pulses because {of the more massive envelope (typically four to five times that of the 1~\Msun{} models).
Also, the metals synthesized during the PIE are more diluted in the more massive envelope.} As a consequence, the opacity does not rise as much as it does in the 1~\Msun{} models and the mass loss is weaker, thus allowing for a ``normal'' thermally pulsing AGB phase.

\subsubsection{A mix of s- and i-processes at [Fe/H]~$=-1.0$}
\label{sect:m2.0z1.0spro}

\begin{figure*}[t]
\centering
\includegraphics[width=1\columnwidth]{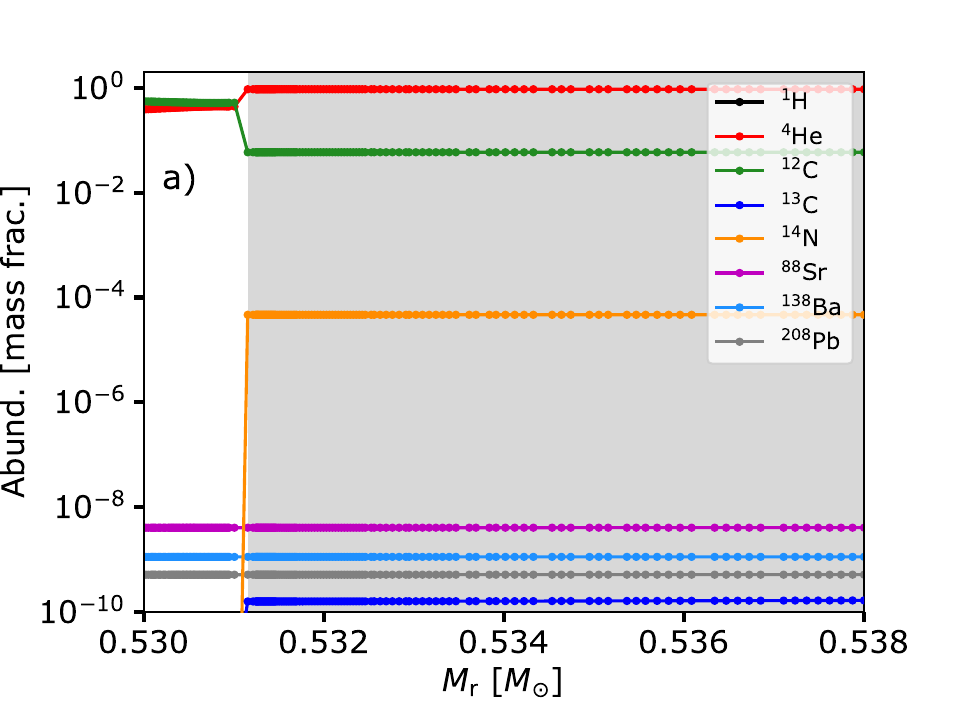}
\includegraphics[width=1\columnwidth]{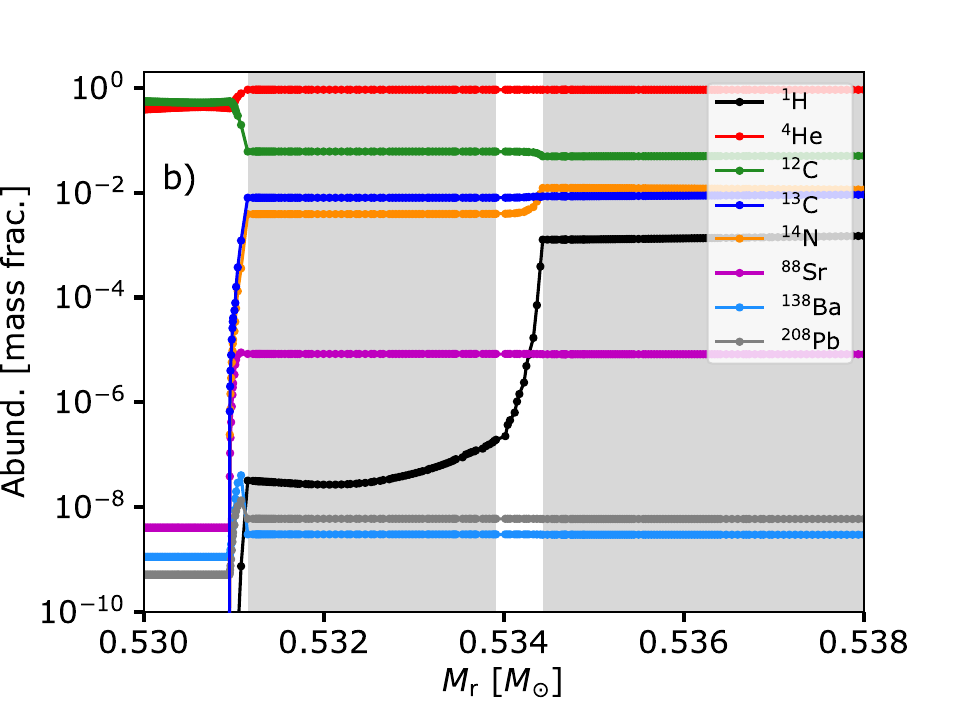}
\includegraphics[width=1\columnwidth]{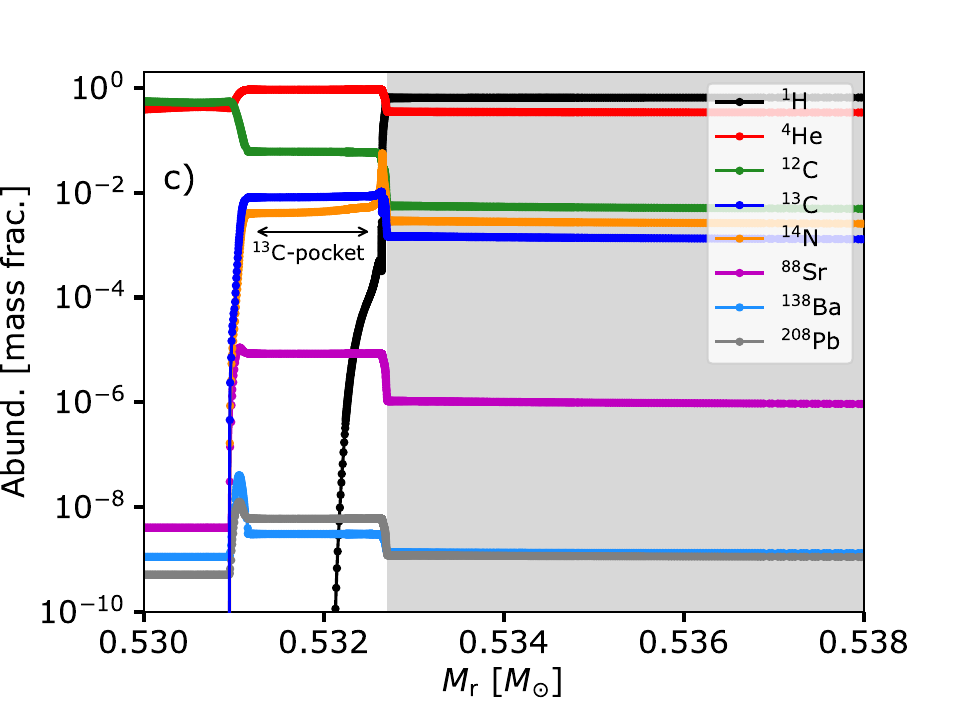}
\includegraphics[width=1\columnwidth]{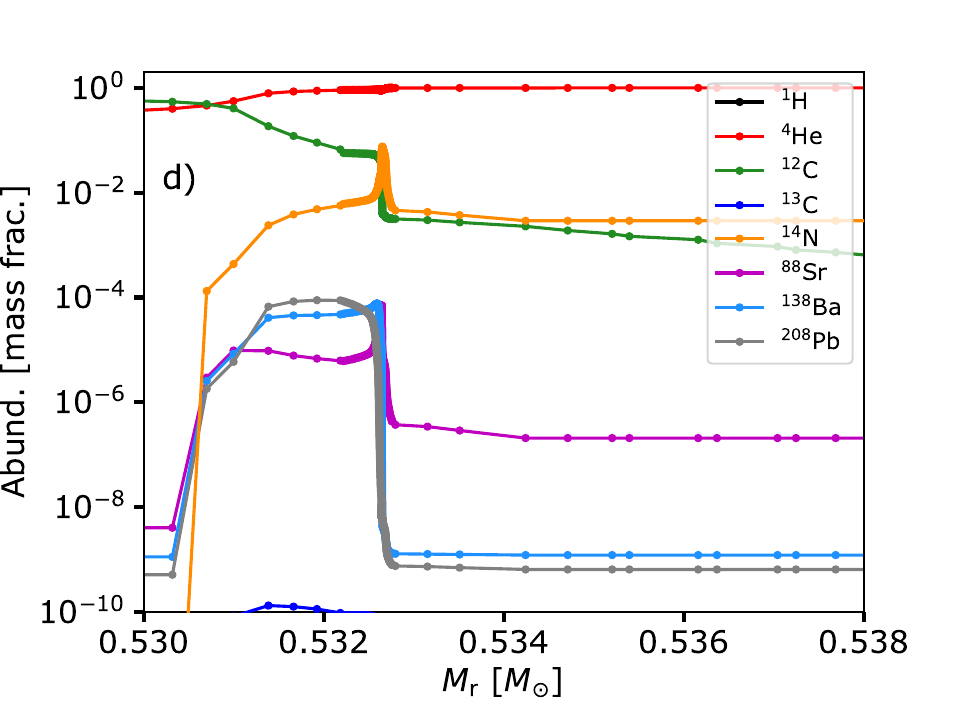}
\includegraphics[width=1\columnwidth]{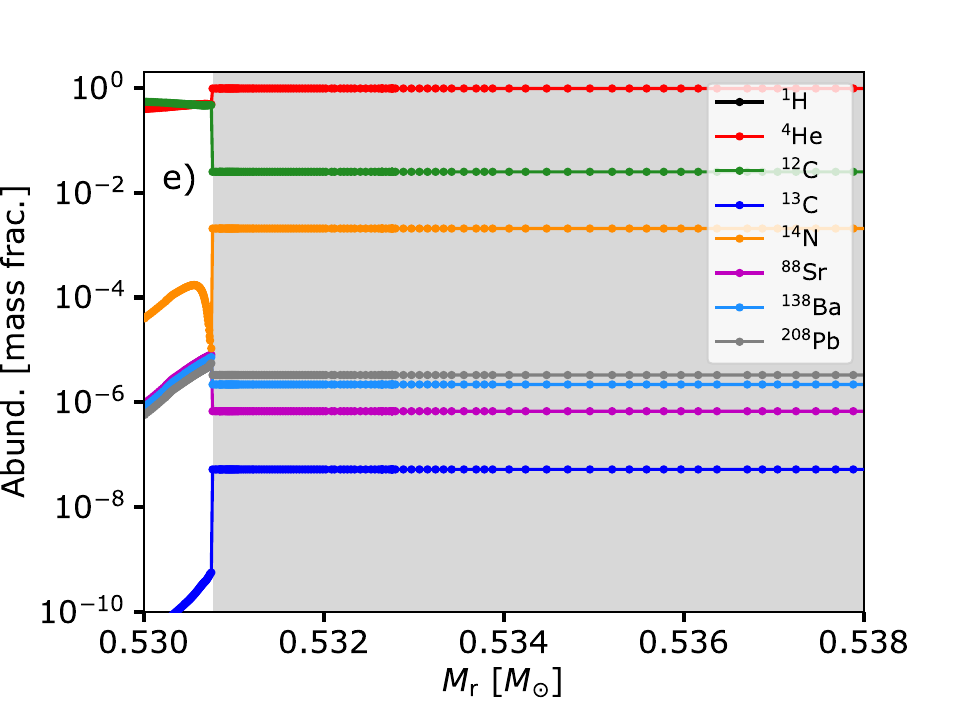}
\includegraphics[width=1\columnwidth]{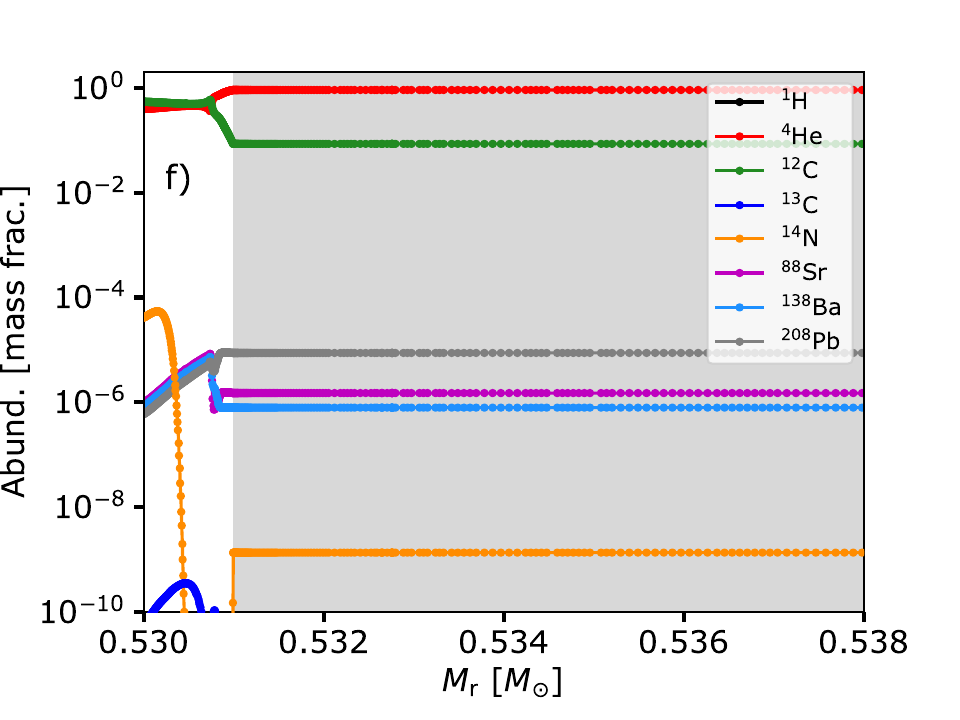}
\caption{Abundance profiles of the M2.0z1.0\_f10 model before the PIE (model 25344, panel a); right after the PIE (model 26044, panel b); just after the merging between the pulse and envelope (model 26544, panel c); at the end of the interpulse (model 42644, panel d); at the very start of the second thermal pulse (model 43586, panel e); and after the second weak PIE (model 44597, panel f).
}
\label{fig:abund_m1.0z2.0}
\end{figure*}

\begin{figure}[t]
\centering
\includegraphics[width=\columnwidth]{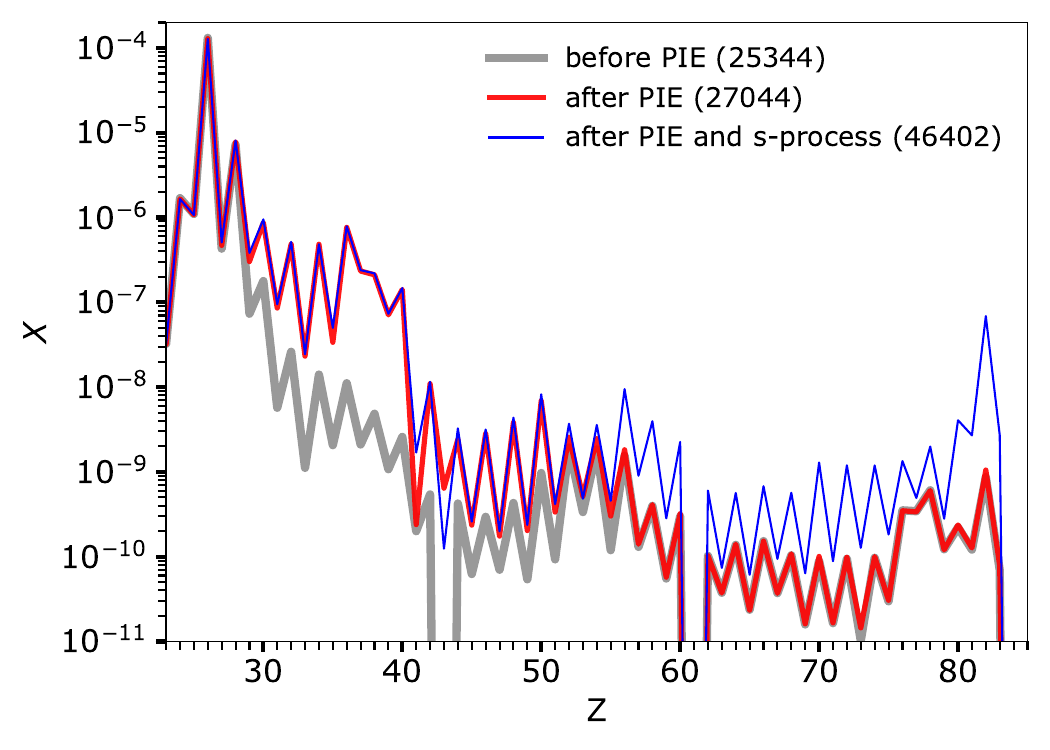}
\caption{Elemental mass fractions at the surface of the M2.0z1.0\_f10 model at three different times. The numbers in parenthesis correspond to the model number of Fig.~\ref{fig:kip_m1.0z2.0}.
}
\label{fig:abund_ipro_spro}
\end{figure}

Our discussion focuses on the evolution of the 2~\Msun{} model at [Fe/H]~$=-1.0$ with $f_{\top} = 0.1,$ which experiences a series of more or less intense PIEs during the first six pulses and a mix of i- and s-processes. The other 2~\Msun{} models with $f_{\top} = 0$, $0.02$ or $0.04$ do not experience any PIE. 
Figure~\ref{fig:kip_m1.0z2.0} shows the internal structure of this model during the first and second pulses.
A PIE starts at model $\sim 25600$, associated with a maximal neutron density of $N_{\rm n, max} = 1.1 \times 10^{14}$~cm$^{-3}$ at  model $\sim 26000$ (Fig.~\ref{fig:kip_m1.0z2.0}, lower panel). 
Shortly after the neutron peak, the pulse splits, which ends the i-process enrichment of the upper part of the pulse (cf. Sect.~3.5 in paper III). 
At this point, both parts of the pulse are enriched in i-process products, especially in the first peak elements such as Sr, as can be seen by comparing panels a and b of Fig.~\ref{fig:abund_m1.0z2.0}. Heavier elements such as Ba and Pb are not significantly produced  because of the high metallicity ([Fe/H]~$=-1$, cf. discussion in Sect.~\ref{sect:nuc1msun}).
At model $\sim 26400$, the pulse merges with the convective envelope (Fig.~\ref{fig:kip_m1.0z2.0}) producing the elemental distribution shown in the red pattern of  Fig.~\ref{fig:abund_ipro_spro}. 
It is important to note that after the merging, hydrogen is dredged down to $M_{r} \sim 0.53$~$M_{\odot}$, producing a significant reduction of the He core mass. As a consequence, the second pulse develops at almost the same mass coordinate as the first one (around $\sim 0.5321$~$M_{\odot}$).

After the split, a radiative zone extending between $M_{r} \sim 0.5312$~\Msun{} and 0.5323~\Msun{} (Fig.~\ref{fig:kip_m1.0z2.0}) with a $^{13}$C/$^{14}$N~$> 1$ (Fig.~\ref{fig:abund_m1.0z2.0}c) forms.
{This `$^{13}$C-pocket' \citep[e.g.,][]{iben82} emerges naturally after the PIE in our models. As explored in various works \citep[e.g.,][]{straniero95,goriely00,busso01, bisterzo10}, it leads to a radiative s-process nucleosynthesis during the interpulse period, which lasts $\sim 5 \times 10^5$~yrs in our model. The temperature of the $^{13}$C-pocket reaches 100~MK and the neutron density goes up to $2.3 \times 10^6$~cm$^{-3}$. 
} 
At the end of this interpulse phase, the abundances of $^{138}$Ba and $^{208}$Pb have increased by $3-4$ dex (Fig.~\ref{fig:abund_m1.0z2.0}d). 
These products are then engulfed in the second thermal pulse (Fig.~\ref{fig:abund_m1.0z2.0}e)
which experiences a weaker PIE (Fig.~\ref{fig:kip_m1.0z2.0}). The neutron density at the bottom of the pulse goes up to $4.5 \times 10^{12}$~cm$^{-3}$ (it stays above $10^{12}$~cm$^{-3}$ for $\sim 0.5$~yr) and barely affects the distribution of heavy elements in the pulse, as can be seen by comparing panels e and f of Fig.~\ref{fig:abund_m1.0z2.0}. 
This second thermal pulse is followed by a third dredge up (Fig.~\ref{fig:kip_m1.0z2.0}) that enriches the surface in s-process products. The net result is an increase of the elements with $55<Z<83$ by typically $\sim 1$ dex (Fig.~\ref{fig:abund_ipro_spro}, blue pattern). 
{The surface chemical composition is not significantly affected by the subsequent evolution. Seven more TPs develop, followed in four cases by very shallow DUPs. In the absence of overshooting below the envelope, there is no more radiative s-process episode. Weak PIEs with maximal neutron densities of $2-5 \times 10^{11}$~cm$^{-3}$ develop during the following pulses but they do not impact the surface abundances. 
The final surface [X/Fe] ratios are shown in the middle panel of Fig.~\ref{fig:abund_xfe} (green distribution).
}

\begin{figure}[t]
\centering
\includegraphics[width=\columnwidth]{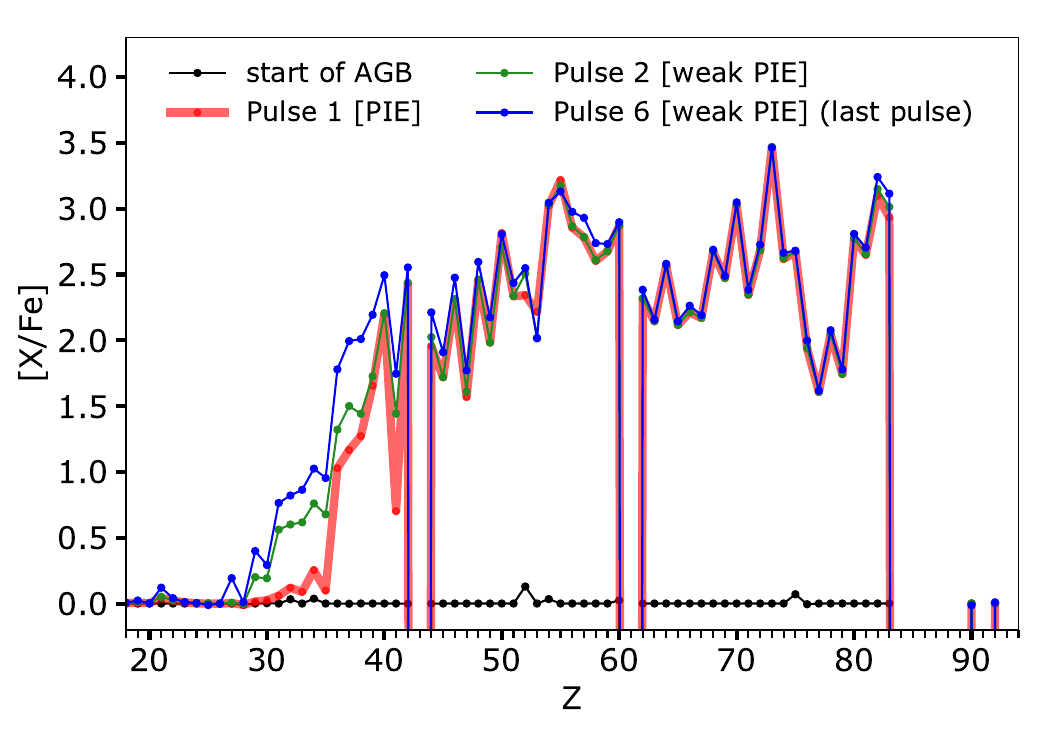}
\caption{Surface [X/Fe] ratios in the M2.0z1.5\_f10 model at four different times. Abundances are shown after the dredge up following the indicated pulse number (except for the black pattern that corresponds to the start of the AGB phase). 
}
\label{fig:abund_xfe_M2.0z1.5}
\end{figure}

\begin{figure}[t]
\centering
\includegraphics[width=\columnwidth]{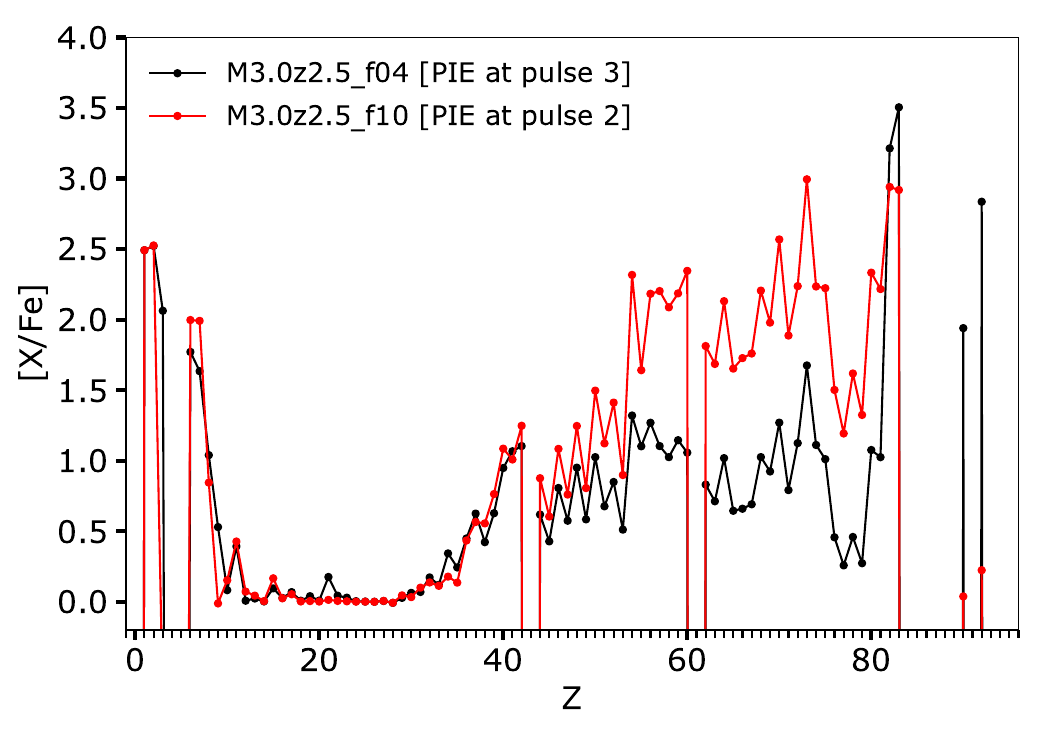}
\caption{Final elemental [X/Fe] ratios at the surface of the M3.0z2.5\_f04 and M3.0z2.5\_f10 models.
}
\label{fig:abund_xfe_3msun_f}
\end{figure}

\subsubsection{Case of a 2~\Msun{} model at [Fe/H]~$=-1.5$}
\label{sect:m2.0z1.5}

The evolution of the 2~\Msun{} model at [Fe/H]~$=-1.5$ with $f_{\top} = 0.1$ resembles that of the 2~\Msun{} model at [Fe/H]~$=-1.0$. A PIE develops during the first pulse,  followed by a radiative s-process episode (with neutron densities of $5 \times 10^{6}$~cm$^{-3}$ at maximum). The s-process products are engulfed in the second pulse, in which a weak PIE takes place. 
The second pulse is then followed by a third dredge up. 
After that, this model experiences five more pulses with weak PIEs (and three DUPs) before it has completely lost its envelope.
 
The resulting nucleosynthesis is however very different compared to the [Fe/H]~$=-1.0$ model (green and blue patterns in Fig.~\ref{fig:abund_xfe}, middle panel). As explained in Sect.~\ref{sect:nuc1msun}, the reduced iron content in the [Fe/H]~$=-1.5$ model allows for a stronger i-process and favors the synthesis of heavier elements. 
The neutron density peak is higher in the [Fe/H]~$=-1.5$ model ($4.7 \times 10^{14}$~cm$^{-3}$) compared to the [Fe/H]~$=-1.0$ model ($1.1 \times 10^{14}$~cm$^{-3}$) resulting in a stronger heavy nuclei production which masks the s-process contribution in the region of $55 < Z < 83$ (cf. the red and green patterns in Fig.~\ref{fig:abund_xfe_M2.0z1.5}).
Ultimately, the chemical yields of this model are mostly determined by the first PIE. Only elements with $30 \lesssim Z \lesssim 40$ are altered (typically 1 dex) by the nucleosynthesis (radiative s-process and weak PIEs) following the first PIE (Fig.~\ref{fig:abund_xfe_M2.0z1.5}).

\subsubsection{Other 2~\Msun{} models}
\label{sect:other2}

The M2.0z0.5\_f10 model follows a very similar evolution as the M2.0z1.0\_f10 model (cf. Sect.~\ref{sect:m2.0z1.0spro}), namely: a weak PIE followed by a radiative s-process which is mixed in the following pulse and eventually dredged up to the surface. 
In \cite{karinkuzhi23}, a 2~\Msun{} model at [Fe/H]~$=-0.5$, with $f_{\top} = 0.06$, $f_{\env} = 0.06$, $D_{\rm min} = 10^7$~cm$^{-3}$ and $p=0.5$ was computed to explain the two newly observed r/s-stars at [Fe/H]~$=-0.5$. This model experienced a PIE during the third thermal pulse, resulting in a rather strong i-process signature. The 2~\Msun{}, [Fe/H]~$=-0.5$ model with $f_{\top}=0.1$ computed in this work also experiences a PIE but during the second pulse instead (Table~\ref{table:3}) and shows a mixed i+s signature (red pattern in middle panel of Fig.~\ref{fig:abund_xfe}). 
The differences in the mixing parameters are likely at the origin of the different chemical patterns. As a matter of fact, the i-process (and potentially s-process) yields remains very sensitive to the overshoot parameters.

A PIE develops in the M2.0z2.0\_f04 and M2.0z2.0\_f10 models during the second and first pulse respectively. The final surface abundances in these two models differ by $\sim 0.5-1$~dex but follows the same trend. In particular, heavy elements such as Pb and Bi are heavily produced (Fig.~\ref{fig:abund_xfe}). In the M2.0z2.0\_f10 model, after the first PIE, the elements with $30 \lesssim Z \lesssim 50$ are slightly enhanced because of the additional weak PIEs and third dredge ups (cf. Sect.~\ref{sect:m2.0z1.5}).

The yields of 1 and 2~\Msun{} models have the same metallicity dependence with a smaller production of heavier elements with increasing metal content.
However, because of its larger mass, the enrichment of PIE products in the 2~\Msun{} models is lower. Also, the s-process signature present in the M2.0z1.0 and M2.0z0.5 models and characterized by the production of elements between $Z \sim 55$ and $Z=83$ is absent in the lower mass models because of their truncated evolution.

\subsection{Evolution and nucleosynthesis of the 3 and 4~\Msun{} models}

The 3~\Msun{} model at [Fe/H]~$=-2$ experiences a PIE during the second pulse for $f_{\top} = 0.1$ followed by five pulses during which weak PIEs develop (like in Fig.~\ref{fig:kip_m1.0z2.0}) with $N_{\rm n, max} = 2.7 \times 10^{12}$~cm$^{-3}$ at maximum. These weak PIEs are sometimes followed by a third dredge up but this barely changes the surface abundances, which are largely determined by the first strong PIE. 
A radiative s-process develops after the PIE of this model but is too weak to significantly alter the i-process signature.

A PIE also develops in the 3~\Msun{} model at [Fe/H]~$=-2.5$, both with $f_{\top} = 0.04$ and $0.10$. The PIE develops during the third (second) pulse for $f_{\top} = 0.04$ ($0.10$).
The pulse conditions are different between the second and third pulses. 
In particular, the maximal temperatures at the bottom of the pulse reach 281 and 266 MK for the $f_{\top} = 0.04$ and $0.10$ cases, respectively. Also, the energy released during these two events are different, which imply different amount of ingested proton.
This impacts the i-process nucleosynthesis and surface abundances (Fig.~\ref{fig:abund_xfe_3msun_f}). 
Like the 3~\Msun{} model at [Fe/H]~$=-2$, weak PIEs develop during the next pulses, altering the surface [X/Fe] ratios of elements with $31<Z<41$ by about 0.3 dex at maximum. Here again, the radiative s-process is too weak to significantly alter the i-process signature.

The global level of enrichment for the 3~\Msun{} is smaller than for the 1 and 2~\Msun{} models since the PIE products are diluted in a larger envelope: at [Fe/H]~$=-2$, the 1, 2 and 3~\Msun{} models have convective envelopes of 0.24, 1.18, and 1.83~\Msun{}, respectively, at the time of the PIE. Also, the pulse mass decreases with increasing mass (0.049, 0.027, and 0.009~\Msun{} for the 1, 2 and 3~\Msun{} models, respectively). To summarize: with increasing stellar masses, the i-process material is diluted in more massive envelopes, resulting in lower level of enrichment but the chemical patterns remain similar (Fig.~\ref{fig:abund_xfe}).

The 4~\Msun{} model at [Fe/H]~$=-2$ does not experience any PIE during the 28 thermal pulses computed, even when adopting $f_{\top} = 0.1$ (Table~\ref{table:3}). At this stage,  only $\sim 0.3$~\Msun\ of envelope is left.  We also confirmed that for high enough $f_{\top}$ values (typically 0.2), a PIE develops during the early AGB phase of this model.

\subsection{Mass and metallicity range of PIEs}

\begin{figure*}[t]
\centering
\includegraphics[width=2\columnwidth]{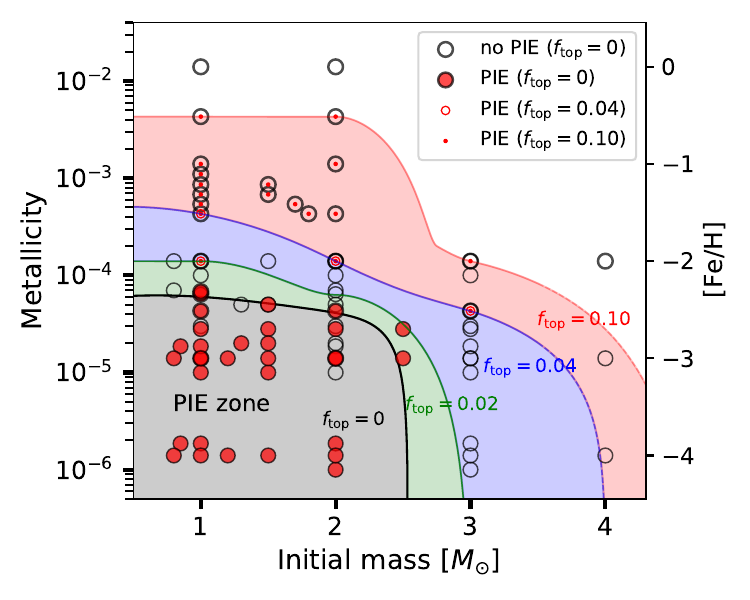}
\caption{
Mass-metallicity diagram showing the occurrence of PIEs during the early AGB phase. The corresponding [Fe/H] ratios are indicated on the right axis assuming solar-scaled mixtures.
Red filled circles show AGB models experiencing a PIE when $f_{\top}=0$. Empty black circles are for models that do not experience a PIE when $f_{\top}=0$. 
Thick black circles denotes models computed in this work and in paper III. The small red circles and red dots highlight our models that experience a PIE when $f_{\top} = 0.04$ and $0.10$, respectively. 
The other models are from \cite{iwamoto04, campbell08, cristallo09a, lau09, suda10, cristallo16}.
The four colored zones show the approximate PIE zone when $f_{\top} = 0$, $0.02$, $0.04$ and $0.10$. 
}
\label{fig:mz}
\end{figure*}

We previously showed that without overshoot, PIEs develop in AGB models having an initial mass below about 2.5~\Msun\ and a metallicity, $Z,$ below about $10^{-4}$ in mass fraction (Paper III). 
This defines a minimal PIE region (or i-process zone), which is represented by the shaded grey area in Fig.~\ref{fig:mz}.
The thick black boundary in Fig.~\ref{fig:mz} is obtained by a classifier using a {Gaussian radial basis function kernel} and trained on both the models from this work and from literature to separate models undergoing PIEs to other models. 

The models computed in this work with different overshoot strengths at the top of the convective pulse ($f_{\top}$) provide a first estimate of the extent of the PIE zone as a function of $f_{\top}$ (Fig.~\ref{fig:mz}). As expected, the higher $f_{\top}$, the more extended this zone. For high enough $f_{\top}$, PIEs can develop close to solar-metallicity. However, it seems that PIE could hardly take place in AGB stars with initial masses higher than 4~\Msun\ (unless the metallicity is very low or $f_{\top}$ very high).

Knowing the location of this PIE boundary is important to assess the contribution of AGB stars to the i-process nucleosynthesis. This boundary remains extremely sensitive to the modeling of overshooting and, in particular, to the adopted value of the $f_{\top}$ parameter.
Hydrodynamical simulations have shown that convection extends beyond the boundary of the convectively unstable region but presently, to our knowledge, there is no clear constraint on the $f_{\top}$ value. 
In particular, we cannot rule out high $f_{\top}$ values of  $f_{\top} = 0.1 $, for instance. 
The $f_{\top}$ values used so far in the literature range between 0.014 and 0.1 (Table~\ref{table:ff}). 
For $f_{\env}$, values higher than 0.1 were sometimes used to obtain a massive enough $^{13}$C-pocket  to account for the surface s-process enrichment in AGB stars \citep[e.g.,][]{pignatari16,ritter18b}.

\begin{figure*}[t]
\centering
\includegraphics[trim = 4cm 0 3cm 0 , width=2\columnwidth]{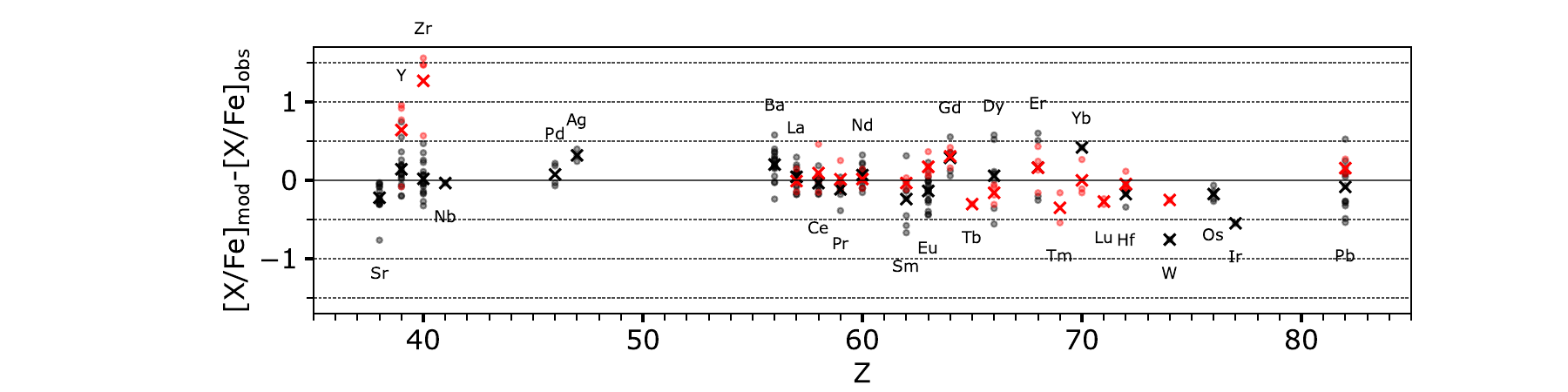}
\caption{Residual of the best fits for the 16 i-process stars candidates of Table~\ref{table:fits}. Black points correspond to MS and RG stars, while red points are for post-AGB stars. The crosses represent the average values of the residuals. The individual fits are shown in Fig.~\ref{fig:fitpagb} and \ref{fig:fits}.
}
\label{fig:res}
\end{figure*}

\begin{figure}[t]
\centering
\includegraphics[width=1\columnwidth]{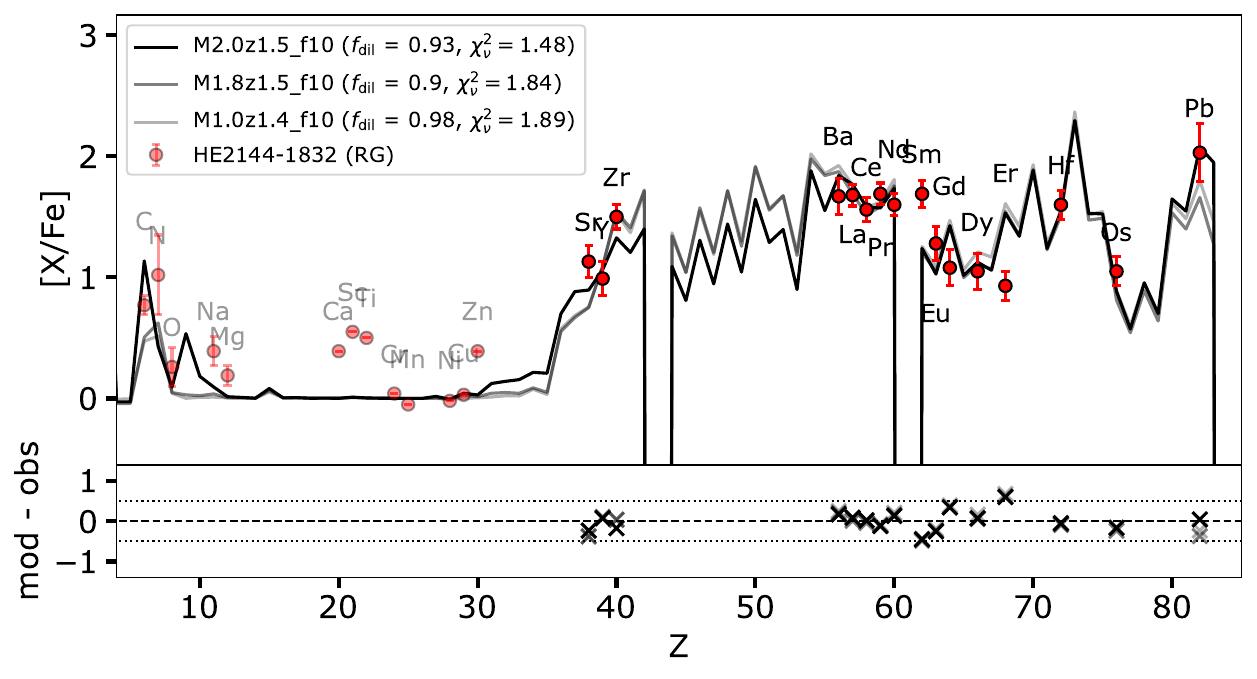}
\includegraphics[width=1\columnwidth]{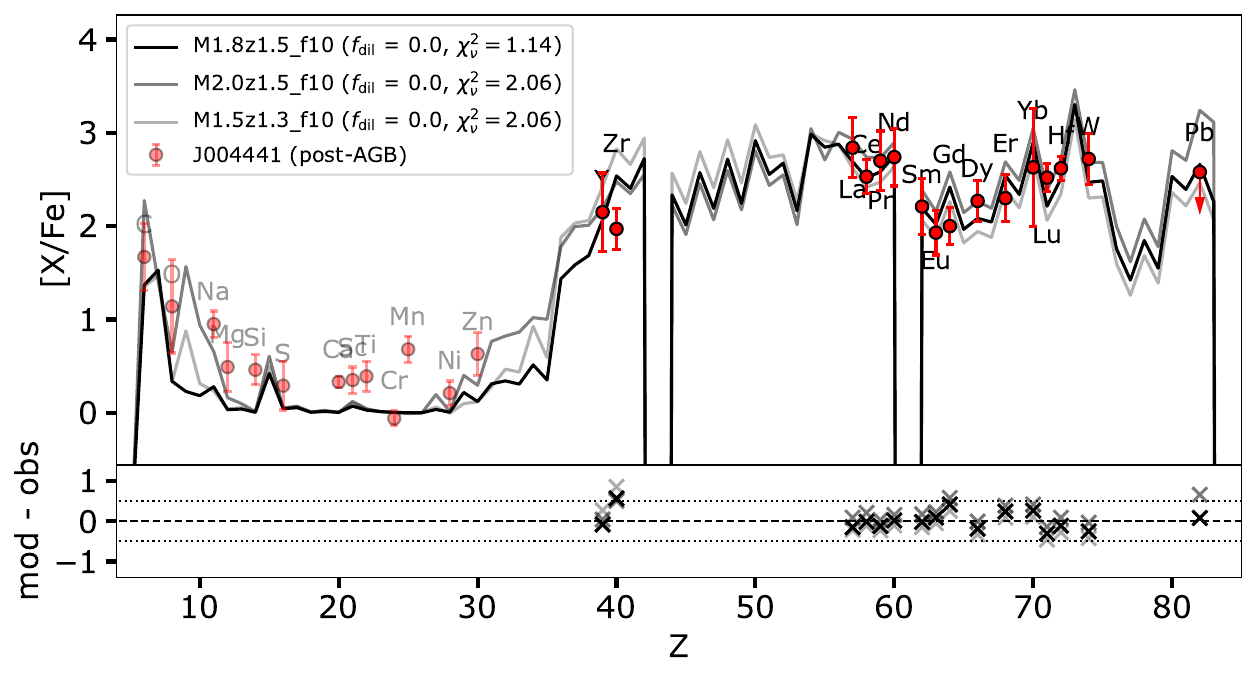}
\caption{{Best description of the RG star HE2144-1832 (top panel)} and post-AGB star J004441 (bottom panel) using the AGB models computed in this work (Table~\ref{table:3}). The three best models are shown in black (lowest $\chi_{\nu}^2$), grey (second lowest $\chi_{\nu}^2$) and light grey (third lowest $\chi_{\nu}^2$). The dilution factor $f_{\rm dil}$ (Eq.~\ref{eq:dilf}, fixed to zero for post-AGB stars) and smallest $\chi_{\nu}^2$ value are indicated. The comparisons for the other stars are shown in Fig.~\ref{fig:fits}. 
}
\label{fig:fitpagb}
\end{figure}

\begin{figure}[t]
\includegraphics[width=1\columnwidth]{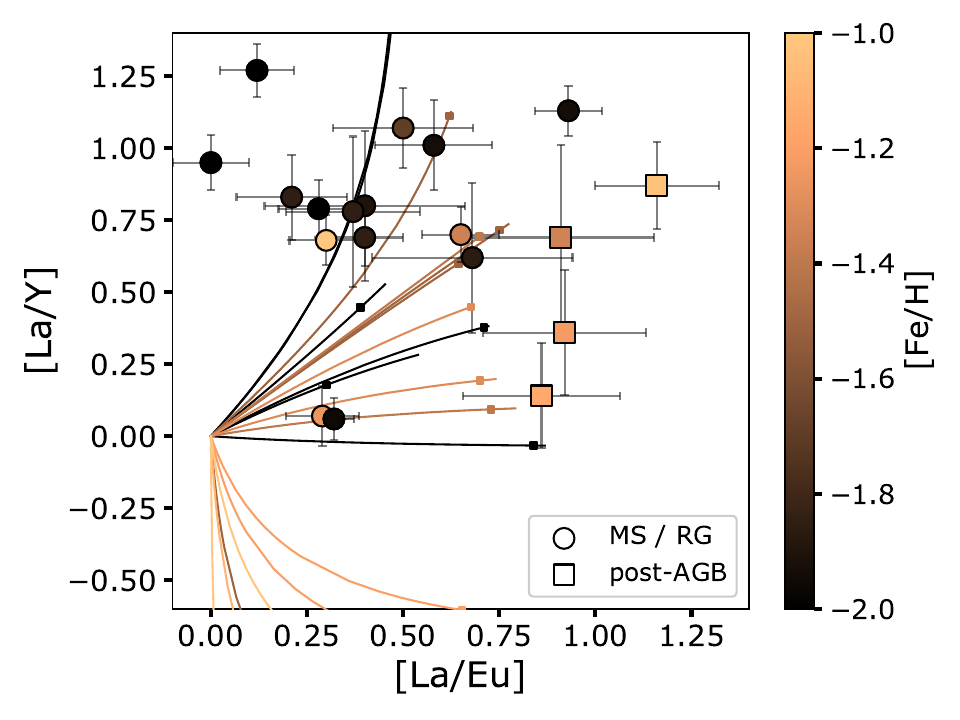}
\includegraphics[width=1\columnwidth]{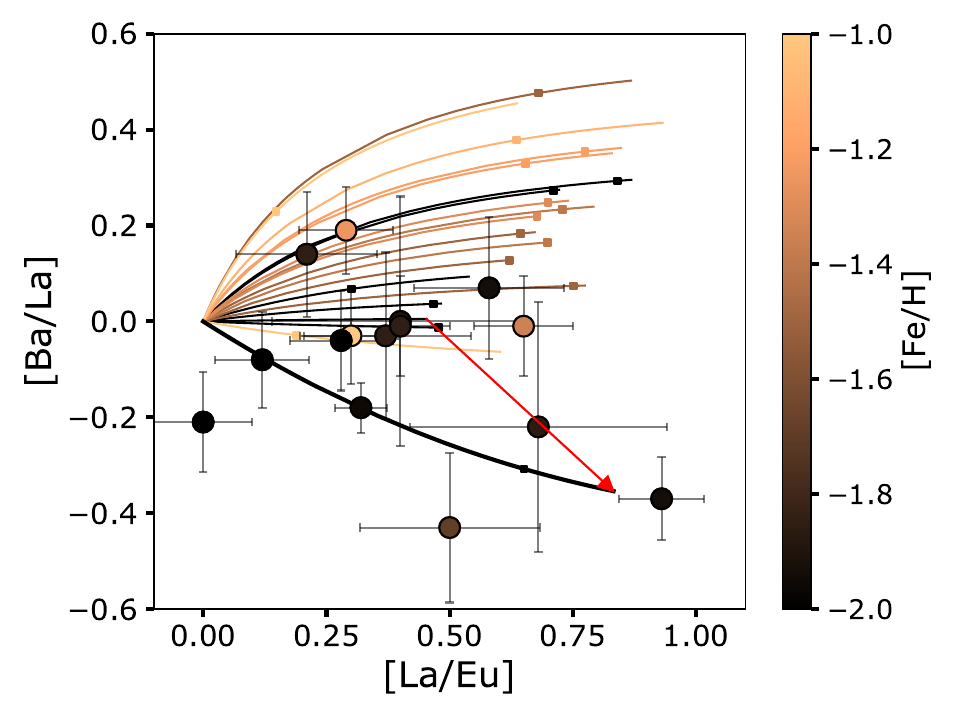}
\caption{
[La/Y] and [Ba/La] abundance ratios as a function of [La/Eu]. 
Circles correspond to MS and RG stars while squares  are for post-AGB stars. 
Lines represent the dilution curves of the AGB material, which ultimately produce a material of solar composition (i.e., ratios equal to zero). 
Lines and symbols are color-coded according to the metallicity [Fe/H].
The small squares on the lines indicates where $f_{\rm dil} = 0.9$.
The thick black line (bottom panel) shows a 2~\Msun{}, [Fe/H]~$=-2$ model computed with a different set of neutron capture rates (see text for more details) and the red arrow shows the resulting abundance displacement. 
}
\label{fig:ratios}
\end{figure}


\section{Comparison with observations}
\label{sect:obs}

In \cite{choplin21} and \cite{choplin22cor}, we have shown that the i-process nucleosynthesis in a 1~\Msun\ AGB model at [Fe/H]~$=-2.5$ was compatible with 14 observed r/s-stars with $-2.70<$~[Fe/H]~$<-2.26$. Here, we examine whether such a comparison remains true when considering higher metallicity stars. We first searched for some r/s-stars candidates with $-2 <$~[Fe/H]~$<-1$ and investigated whether our AGB models with overshoot can account for their chemical abundance patterns.

\subsection{r/s-stars sample}
\label{sect:subsect_selec}

To search for r/s-stars candidates, we used the $d_{\rm RMS}$ proxy introduced in \cite{karinkuzhi21} to classify s-, r-, and r/s-stars. It is defined as
\begin{equation}
d_{\rm RMS} = \left( \frac{1}{N} \sum_{i=1}^{N} (A_{ i,\star} - A_{i,r})^2 \right)^{1/2}
\label{eq:drms}
,\end{equation}
with $N$ as the number of considered heavy elements and $A_{ i,\star} = \log_{10}(n_{i,\star}/n_{\mathrm{H},\star}) + 12,$ where $n_{i,\star}$ is the number density of element $i$ in the sample star.  The quantity $A_{i,r}$ corresponds to the solar r-process abundance of element $i$ scaled to the Eu abundance of the sample star. It is computed as
\begin{equation}
A_{i,r} = A_{i,r,\odot} - (A_{\mathrm{Eu},r,\odot} - A_{\mathrm{Eu},\star}),
\label{Air}
\end{equation}
where $A_{i,r,\odot}$ is the solar r-process abundance of element $i$ from \cite{arnould07}. 
The s-, r-, and r/s-stars have been shown to be rather well identified when using this criterion \citep[Fig.~10 of][]{karinkuzhi21}.
The r/s-stars are characterized by intermediate $d_{\rm RMS}$ values, typically between $0.5$ and $1$. 
For our study, we selected {potential i-process stars candidates} from the SAGA database \citep{suda08, suda17}. We complemented these data with a few recent observations (see Table~\ref{table:fits} for more details) using the following filters: (1) a metallicity in the range $-2 <$~[Fe/H]~$<-1$, (2) an enrichment in barium relative to iron [Ba/Fe] $> 0.5$, (3) at least five measured abundances between Ga ($Z=31$) and Pb ($Z=82$), and (4) a heavy-element pattern characterized by $0.4<d_{\rm RMS}<1$.
This gives us a sample of 22 stars: 8 dwarfs or Main Sequence (MS), 10 red giants (RG), and 4 post-AGB stars, as summarized in Table~\ref{table:fits}.

\subsection{Different scenarios for MS and RG stars and post-AGB stars}
\label{sect:scenar}

For MS and RG stars, the enrichment in trans-iron elements {could come from a now-extinct AGB companion} that polluted the secondary through winds \citep[e.g.,][]{abate13}. This scenario predicts that r/s-stars ought to reside in binary systems. This was indeed shown in most cases \citep[e.g., 9 out of 11 r/s-stars in][are confirmed binaries]{karinkuzhi21}. {In our sample however, only 6 MS/RG stars out of 18 were clearly} identified as binaries (Table~\ref{table:fits}), while the 9 other stars are either single, {long-period binary systems or binary systems with a high orbital axis inclination, which hinders the detection of radial velocity variations. The binarity of the remaining three RG stars was not investigated, to our knowledge. }
{Radial velocity monitoring over long periods of time is desired to unveil the binary status of these stars. }
{An alternative scenario that does not require binarity is to rely on an early generation of AGB stars that polluted with i-process material the natal cloud of these stars.}

The four post-AGB stars in our sample may be intrinsically enriched, having synthesized heavy element in their interior. In this case, there is no need to account for an AGB companion as it would be for MS and RG stars. The surface composition can be directly compared to our AGB model predictions. 
{To do so, we assumed that the final surface abundances of our AGB models are not further affected by the late AGB and post-AGB evolution and, therefore, they do reflect the post-AGB abundances. Although most of our models have reached the end of the AGB phase (Table~\ref{table:3}), they may still experience a late pulse, possibly altering their final surface abundances \citep{herwig11,desmedt12}. Computing the post-AGB phase of these models is beyond the scope of this work, but this would be required to strengthen our comparisons with observations.}

\begin{table*}[t]
\scriptsize{
\caption{
{Characteristics and adopted dilution parameters of the 22 selected stars. Their class is indicated by MS (main sequence), RG (red giant) and post-AGB as well as their metallicity, approximate mass (taken from the literature) and binary status: "B" for detected binary and 'S/L/I' for either single ("S"), long-period binary ("L") or binary systems with a high orbital axis inclination ("I"). The quantity $N_{\rm ab}$ is the number of elements observed between Zn ($Z=30$) and Bi ($Z=83$), $d_{\rm RMS}$ the average distance to the solar scaled r-process (Eq.~\ref{eq:drms}), $\chi_{\rm \nu,min}^2$ the minimum reduced $\chi_{\nu}^2$ value, $f_{\rm dil}$ the dilution factor (Eq.~\ref{eq:dilf}), $M_{\rm acc}$  the amount of mass the observed star must accrete (in the binary case) to reproduce its current level of enrichment (as defined in Sect.~\ref{sect:accms}).} 
\label{table:fits}
}
\begin{center}
\resizebox{18.5cm}{!} {
\begin{tabular}{lllccllllcl}
\hline
 Star          & Class   & [Fe/H] & $M_{\rm ini}$ [M$_{\odot}$] & Bin & $N_{\rm ab}$ & $d_{\rm RMS}$ & $\chi_{\rm \nu,min}^2$ & $f_{\rm dil}$ & $M_{\rm acc}$ [M$_{\odot}$] &  Ref.                                 \\
\hline
 BS16080-175   & MS   & -1.86  & ? & S/L/I & 7            & 0.96           & 0.72                   & 0.92          & $4.3 \times 10^{-5}$  & 1, 2                                 \\
 BS17436-058   & RG   & -1.90   & ? & S/L/I & 7            & 0.73           & 0.6                    & 0.93          &$3.0 \times 10^{-2}$  & 1, 2                                  \\
 CS22880-074   & MS   & -1.93  & ? & S/L/I & 11           & 0.95           & 0.65                   & 0.995      &$2.5 \times 10^{-6}$    & $2-5$                         \\
 CS22887-048   & MS   & -1.85  & ? & B    & 7            & 0.70           & 0.27                   & 0.45      &   $6.1 \times 10^{-4}$ & 1, 6                            \\
CS29503-010  & MS    & -1.69  & $0.86$ & S/L/I & 10           & 0.74           & 2.54                   & 0.10          &$4.5 \times 10^{-3}$  & 1, 2, 7                            \\
 CS29513-032  & MS    & -1.85  & ? & S/L/I & 12           & 0.65           & 1.56                   & 0.99        &$5.1 \times 10^{-6}$   & 8                               \\
 HD106038     & MS    & -1.48  & $0.7$ & S/L/I & 7            & 0.79           & 1.22                   & 0.992       &$4.0 \times 10^{-6}$  & $9 -14$            \\
 HD126681     & MS    & -1.17  &  $0.7$ & S/L/I & 8            & 0.60           & 0.84                   & 0.996     &  $2.0 \times 10^{-6}$  & 8, $14-20$   \\
 HD166161     & RG    & -1.25  & ? & ? & 12           & 0.83           & 0.68                   & 0.9997    &$1.2 \times 10^{-4}$    & 13, $21-28$    \\
 HD166913     & MS    & -1.93  & $0.73$ & S/L/I & 6            & 0.56           & 0.48                   & 0.93        & $3.8 \times 10^{-5}$  & 9, 14, 15, 29, 30               \\
 {HD206983}    & {RG}    & -1.00  & ? & ? &      11       &      0.68      &          0.36      &    0.999     & $4.0 \times 10^{-4}$  &    31, 33            \\
 {HD209621}    & {RG}    & -2.00  & ?  & B &      12       &       0.52     &        4.32 (2.46$^*$)      &   0.85  & $7.1 \times 10^{-2}$ &     31, 34           \\
 {HD5223}    & {RG}    & -2.00  & ?  & B  &     14        &        0.63    &       1.93      &   0.95      &$2.1 \times 10^{-2}$  &     31, 34, 35           \\
 {HE0507-1653}    & {RG}    & -1.35  & ?  & B  &       16      &   0.87         &     2.51       &  0.91    & $4.0 \times 10^{-2}$ &  31, 32, 36              \\
 {HE1120-2122}    & {RG}    & -2.00  &  ? & B  &       16      &    0.44     &      2.19 (1.25$^*$)    & 0.95  & $2.1 \times 10^{-2}$ &      31, 32          \\
 {HE2144-1832}     & {RG}    & -1.85  & ? & B &      16       & 0.65           &   1.48           &  0.93   & $3.0 \times 10^{-2}$ &       31, 32         \\
 J130200.0-084328 & RG & -1.95  & ? & S/L/I & 10           & 0.80           & 1.61                   & 0.72        &$1.6 \times 10^{-1}$    & 37                             \\
 T6953-00510-1   & RG & -1.93  & ? & ? & 7            & 0.97          & 0.74                   & 0.997       &$1.2 \times 10^{-3}$    & 38                              \\
\hline
 J053250      & post-AGB    & -1.22  & $1 - 1.5$ & ? & 13           & 0.99           & 6.60 (1.13$^*$)             & 0.0        &$-$   & 39                              \\
 J052043    &  post-AGB     & -1.15  & $1 - 1.5$ & ? & 14           & 0.97           & 5.66  (0.51$^*$)            & 0.0        &$-$   & 39, 40                              \\
 J051848      &  post-AGB    & -1.03  & $1 - 1.5$ & ? & 14           & 0.93           & 6.44 (1.66$^*$)          & 0.0        &$-$   & 41                              \\
 J004441      &  post-AGB    & -1.34  & $1.3$& ? & 15           & 0.81           & 1.14             & 0.0        &$-$    & 40, 42                              \\
\hline
\end{tabular}
}
\end{center}
}
\normalsize{
{References}. 1 - \cite{allen12}; 2 - \cite{tsangarides05}; 3 - \cite{aoki02b}; 4 - \cite{preston01}; 5 - \cite{aoki02c}; 6 - \cite{masseron12}; 7 - \cite{aoki07}; 8 - \cite{roederer14a}; 9 - \cite{tan09}; 10 - \cite{boesgaard11}; 11 - \cite{fabbian09}; 12 - \cite{gratton03}; 13 - \cite{nissen07}; 14- \cite{hansen12}; 15 - \cite{melendez10}; 16 - \cite{bensby14}; 17 - \cite{caffau05}; 18 - \cite{reddy06}; 19 - \cite{nissen11}; 20 - \cite{yan16}; 21 - \cite{gratton00}; 22 - \cite{simmerer04}; 23 - \cite{fulbright03}; 24 - \cite{mishenina01}; 25 - \cite{takeda11}; 26 - \cite{mishenina02}; 27 - \cite{roederer10}; 28 - \cite{burris00}; 29 - \cite{jonsell05}; 30 - \cite{bihain04}; 31 - \cite{karinkuzhi21}; 32 - \cite{jorissen16}; 33 - \cite{masseron10}; 34 - \cite{mcclure90}; 35 - \cite{goswami06}; 36 - \cite{hansen16a}; 37 - \cite{sakari18}; 38 - \cite{ruchti11}; 39 - \cite{vanaarle13}; 40 - \cite{desmedt14}; 41 - \cite{desmedt15}; 42 - \cite{desmedt12}

$^*$ The numbers in parenthesis indicate the $\chi_{\rm \nu,min}^2$ values {when excluding Sr, Y and Zr from the adjustment}.
}
\end{table*}

\subsection{Fitting procedure}
\label{sect:fitproc}

We followed the same procedure as in Sect.~6.2 of \cite{choplin21} to find the lowest $\chi^2$ among our AGB models. 
In particular, we used $\chi_{\nu}^2 = \chi^2 / N_{\rm ab}$, which represents the $\chi^2$ normalized by the number of data points (i.e., by the number of derived elemental abundances). When the AGB material is diluted in the unpolluted envelope of the MS or RG companion ($X_{\rm ini}$), the resulting mass fraction of an isotope $i$ is given by: 
\begin{equation}
X_{i} =  (1-f_{\rm dil}) \, X_{s} + f_{\rm dil} \, X_{\rm ini}
\label{eq:dilf}
,\end{equation}
where $0 \leq f_{\rm dil}<1$ is the dilution factor, while $X_{s}$ and $X_{\rm ini}$  are the surface and initial mass fractions of isotope $i$, respectively. For post-AGB stars, we have $f_{\rm dil} = 0$.

To compute the $\chi_{\nu}^2$ value, we consider the abundance of elements heavier than Zn ($Z=30$). Nuclei from Na ($Z=11$) to Zn ($Z=30$) are scarcely impacted by low-mass AGB nucleosynthesis and their presence in the stellar envelope may originate from previous sources that polluted the proto-stellar gas (e.g., winds and/or core-collapse supernovae of massive stars). 
The C, N, and O abundances are more difficult to interpret as these elements are impacted by the AGB donor nucleosynthesis, could be present in non-solar proportions in the proto-stellar cloud or be altered by internal mixing processes in the observed star when it evolved, for example, into a giant.
These elements are discussed in the next sections, but they are not considered in the determination of the minimal $\chi^{2}_{\nu}$.

For each of the 22 observed stars, we selected the three best AGB models for which the metallicity is within 0.5 dex of the observed value (e.g., only our models with [Fe/H]~$=-1.5$ and $-2.0$ are considered to describe the abundances of BS16080-175, which has a metallicity of [Fe/H]~$=-1.86$).

\subsection{Heavy elements}
\label{sect:resfit}

\subsubsection{MS and RG stars}

A reasonable agreement between models and observations was found {for the 18 MS and RG stars} (Figs.~\ref{fig:res}, \ref{fig:fitpagb}, \ref{fig:fits}), with residuals less than  $\pm 0.5$~dex (Fig.~\ref{fig:res}) {in most cases}. 
This is reasonable in view of the various uncertainties associated with observations (typically $0.2-0.5$~dex), numerics \citep[$\pm 0.3$~dex on the abundances][]{choplin21}, nuclear physics \citep[e.g.,][]{goriely21, martinet23,choplin22a}, and mixing processes, such as the overshoot description discussed in Sect.~\ref{sect:allparam}. 

As shown in Fig.~\ref{fig:ratios}, the different abundance ratios can be reasonably well accounted for by our models. {Two stars (HD209621 and HE1120-2122) have $\rm{[La/Eu]}\simeq 0$ and [La/Y]~$\simeq 1,$ which is $\simeq 0.5$~dex away from the closest model track. 
These stars have a particularly low [Y/Fe] (about 0.5, Fig.~\ref{fig:fits}) but high abundances of Sr and Zr which are nearby elements. This scatter cannot be reproduced by our models, which predict minimal $\chi_{\nu}^2$ values of 4.32 and 2.19 (Table~\ref{table:fits}, these values drop to 2.46 and 1.25 when excluding Sr, Y and Zr from the adjustment)}. 
One RG star (T6953-00510-1) has [La/Eu]~$=0.93,$ which is at the limit of our model predictions ($\sim 0.8$) and exhibits a value of a $d_{\rm RMS} = 0.97$  (Table~\ref{table:fits}), which may point towards an s-process (rather than i-process) as its origin. 
{Although the negative [Ba/La] ratio of several stars is not compatible with our model predictions (Fig.~\ref{fig:ratios}, bottom panel), \cite{martinet23} showed, based on a 1~\Msun{} AGB model with [Fe/H]~$=-2.5$, that nuclear uncertainties introduce a spread in the [Ba/La] ratio of $-0.75< \mathrm{[Ba/La]} < 0.63$ (their Sect.~3.3.1).} As a test, we re-computed the PIE in the M2.0z2.0\_f10 model using a different set of ($n,\gamma$) rates and found that a final surface [Ba/La] ratio lower by about 0.4 dex, giving a dilution curve relatively consistent with observations with [Ba/La]~$<0$. 
It remains to be checked if the observational scatter in Fig.~\ref{fig:ratios} can be covered when this alternative set of nuclear rates is adopted for all our stellar models.
This task goes beyond the scope of the present paper but would be required to draw solid conclusions.

\subsubsection{Post-AGB stars}

For post-AGB stars, there is no adjusting dilution factor and good-quality fits are more difficult to obtain. 
Nevertheless, the post-AGB star J004441 can be well explained by our M1.8z1.5\_f10 AGB model (Fig.~\ref{fig:fitpagb}, bottom panel). 
This agreement is also acceptable for the three other post-AGB stars except for Y and Zr, which are overestimated in our calculations (Fig.~\ref{fig:res} and \ref{fig:fits}). For these objects, $\chi_{\rm \nu,min}^2 \sim 5-6$ but if we exclude Y and Zr from the fit, it drops to $0.5 - 1.7$ (Table~\ref{table:fits}).
As for the RG star T6953-00510-1, the [La/Eu] ratio of post-AGB stars is hard to reconcile with our predictions. The post-AGB J051848 has the highest value of [La/Eu]~$=1.16$ which is $\sim 0.4$~dex above our model values (Fig.~\ref{fig:ratios}), but compatible with nuclear uncertainties.
The rather high [La/Eu] ratios together with the $d_{\rm RMS}$ of $0.8 - 1$ (Table~\ref{table:fits}) indicate that these objects have an unclear chemical signature between the s- and the i-process.

\subsection{The C, N, O elements}
\label{sect:cno}

{The [C/Fe] ratios can be relatively well accounted for (Figs.~\ref{fig:fitpagb} and \ref{fig:fits}), even if they were not included in the $\chi_{\nu}^2$ fitting procedure. 
The exception is CS22880-074 with [C/Fe]~$=1.42$ while our best models predict [C/Fe]~$\leq 0.5$. Three other stars (BS160080-175, BS17436-058 and HD5223) have a [C/Fe] ratio $\simeq 0.5$~dex above our model predictions.
}
{The agreement for [N/Fe] is acceptable for several stars. However, for five stars -- BS17436-058, HD206983, HD209621, HE1120-2122, and T6953-00510-1 -- [N/Fe] is $\sim 1$~dex above our model predictions ($\sim 2$~dex for T6953-00510-1).}
{Finally, the high [O/Fe] of $\simeq 1$ of several MS and RG stars -- HD126681, HD166161, HD166913, HD209621, HE1120-2122 and T6953-00510-1 -- cannot be accounted for by our models. 
The four post-AGB stars have also [O/Fe]~$\simeq 1,$ but our models predict [O/Fe]~$\simeq 0.5$ at maximum.}

{In summary, for the stars mentioned above, C, N, and O are always underproduced by our models, which may suggest an early CNO enrichment by external sources such as massive stars.
Some of the discrepancies for [C/Fe] and [O/Fe]} can be attributed to the fact that we did not consider $\alpha$-enrichment in our initial composition. {In particular}, with [O/Fe] ratios increased by 0.4-0.7 dex for metallicities below [Fe/H]~$ < -1$ \citep{Bensby2014}, the agreement would be significantly improved.  
Nevertheless, using a different initial mixture may potentially impact the PIEs because the metallicity ($Z$) is affected, thereby preventing us from drawing firm conclusions without computing additional models.
{The high N abundances might originate} from previous rotating massive stars that are known to overproduce N, especially at low metallicities \citep[e.g.,][]{meynet06}.

\subsection{Accretion and dilution for MS and RG stars}
\label{sect:accms}

{Although the binary status is not confirmed for a significant fraction of the sample stars (Sect.~\ref{sect:scenar} and Table~\ref{table:fits}), we can still estimate how much mass the r/s-stars should have accreted from their AGB companion, if they were all indeed binary.}
As developed in \cite{choplin21,choplin22cor}, in the binary mass transfer scenario, the dilution factor $f_{\rm dil}$ (Eq.~\ref{eq:dilf}) can be linked to the envelope mass of the r/s-star before the accretion episode ($M_{\env}$) and the mass accreted by the r/s-star ($M_{\rm acc}$) as:
\begin{equation}
f_{\rm dil} = \frac{M_{\env}}{M_{\env} + M_{\rm acc}}.
\label{eq:dilfac2}
\end{equation} 
For a low-metallicity 1~\Msun{} star, the envelope mass on the main sequence is on the order of $M_{\env} = 5 \times 10^{-4}$~\Msun\  and around $M_{\env} = 0.4$~\Msun\ on the giant branch. Using these values, {we find (Table~\ref{table:fits}) $2.0 \times 10^{-6} < M_{\rm acc}/ M_{\odot} < 0.16$, with 11 out of 18 stars having $M_{\rm acc} < 10^{-2}$~$M_{\odot}$ and 9 out of 18 star with $M_{\rm acc} < 10^{-3}$~$M_{\odot}$. }
The star with the highest $M_{\rm acc}$ of $0.16$~\Msun\ is J130200.0-084328. Its best fit is obtained by our 3\Msun\ AGB model at [Fe/H]~$=-2.0$ (Fig.~\ref{fig:fits}). Considering that more than 2~\Msun\ will be lost by the AGB phase of this model, it does not seem unrealistic to assume that the secondary could accrete 0.16 \Msun \ (i.e., $\sim 8\%)$ of that wind material. 
Although simple, this estimate confirms that the  accretion scenario for these r/s-stars is not unrealistic.
However, as discussed in Sect.~\ref{sect:fitproc}, the binary status of our sample stars remains unclear and will need to be elucidated.

\subsection{Comparison with previous studies: r+s scenario}
\label{sect:compaother}

{Eight of our sample stars, BS16080-175, BS17436-058, CS22880-074, CS22887-048, CS29513-032, HD206983, HD209621, and HD5223, were analyzed by \cite{bisterzo12}. They used AGB models computed by \cite{bisterzo10}, which experience an s-process nucleosynthesis through the artificial introduction of a $^{13}$C-pocket with varying efficiencies.
In seven out of these eight stars, they considered an r+s scenario: they mixed a variable fraction of solar r-process material to the s-process material of their AGB models to fit the observed abundances. The abundances of these stars are reasonably well reproduced in the framework of the r+s scenario, except for Y and W in HD209621 (overproduced and underproduced by $1 - 1.5$~dex in the model) and Y in HD5223 (overproduced by $0.5 - 1$~dex). We faced rather similar issues for these stars, although the agreement with W in HD209621 is relatively better (underproduced in our models by 0.7 dex, Fig.~\ref{fig:fits}).
These authors did not consider an r-process contribution for the eighth star (CS22880-074), but its abundances are reasonably well reproduced by an s-process operating in the 1.3~$M_{\odot}$ AGB model (except for Er which is underestimated by 0.5~dex).  
}
{The observed abundances of the post-AGB star J004441 were also shown to be well reproduced in the framework of the r+s scenario \cite{cui14}}. 
{At this stage, there is no clear preference between the i and r+s scenario for these stars. 
A dedicated study would be required to determine whether certain key observations (especially elemental or isotopic ratios) could help in distinguishing between the i- versus the r+s scenarios. 
}


\section{Summary and conclusions}
\label{sect:concl}

In this work, we studied the i-process developing during a PIE in AGB stellar models of various initial masses and metallicities, including the possibility to have overshoot mixing at convective boundaries. 
The models were computed with the code STAREVOL in which a network of 1160 nuclei is used and coupled to the transport equations. 

A detailed study of the impact of all overshoot parameters was first carried out during the PIE of a 1~\Msun{}, [Fe/H]~$=-2.5$ model. In particular, we considered different overshoot coefficients above and below the thermal pulse, as well as below the convective envelope.  
While a PIE always develops in this model star regardless of the overshoot assumptions, the final surface abundances of $36<Z<56$ elements show an overall scatter of $0.5-1$~dex, which goes up to $2-3$~dex for Th and U.
Because the abundances of $56<Z<80$ elements are impacted by less than 0.5~dex by the different overshoot assumptions (at least in the considered AGB model), the predictive power of our i-process models is higher for these nuclei. 
Actinides are only significantly produced if the overshoot at the top of the convective pulse is low enough ($f_{\top} \lesssim 0.04$).

We then investigated AGB models with $1 \leq M_{\rm ini}/M_{\odot}  \leq 4$ and metallicities $-2\leq$~[Fe/H]~$\leq 0$. In these cases, the overshoot mixing at the top of the convective thermal pulse is found to play a key role in favoring the development of a PIE. 
While for $f_{\top} = 0$, no PIE develops, for low (0.02), medium (0.04), and high (0.10) $f_{\top}$ values, we found that 6~\%, 24~\%, and 86~\% of our AGB models, respectively, experience a PIE (Table~\ref{table:3}). 
In the mass-metallicity diagram, the PIE region is extended with increasing $f_{\top}$, almost reaching solar metallicity for high $f_{\top}$ values (Fig.~\ref{fig:mz})
In the framework of our calculations, the chemical imprint of the i-process is increased with decreasing metallicity, and mainly affects heavy elements with $Z > 30$. 
We also found that PIEs leave a $^{13}$C-pocket at the bottom of the pulse that can give rise to an additional radiative s-process nucleosynthesis. The s-process products either stay locked deep into the star or are dredged-up to the surface but, in this case, remain generally  overwhelmed by the higher i-process enrichment. In our 2~\Msun{} models at [Fe/H]~$=-1.0$ and $-0.5$, this $^{13}$C-pocket leads to a noticeable mixed signature of i+s elements at the AGB surface.
After the first main PIE, our models with $M_{\rm ini}  > 1$~$M_{\odot}$ experience weak PIEs that further impact the surface abundances of $31<Z<41$ elements. 

Finally, based on the classification scheme of r/s-stars introduced by \citet{karinkuzhi21}, {a sample of 22} observed main sequence, giant, and post-AGB stars with $-2<$~[Fe/H]~$<-1$ was selected and compared to our models. The [X/Fe] ratios of nuclei with $Z>30$ can be reasonably well reproduced by our models, with residuals {generally restricted to} the range of $\pm 0.5$~dex, {with a clear exception for} Y and Zr in three post-AGB stars. 
{The overall good agreement found between r/s-stars candidates and our AGB models at $-2<$~[Fe/H]~$<-1$ is in favour of an i-process operating in AGB stars up to [Fe/H]~$\simeq -1$ at least.}
{Radial velocity monitoring over long periods of time is desired to unveil the binary status of these r/s-stars.}
{If these stars are in binary systems,} our simple estimate for the mass that should be accreted 
from the AGB companion to explain the level of i-process enrichment leads to realistic values. 
{If these stars are single, an alternative scenario needs to be considered, for instance, by assuming they formed from an i-process enriched material left by an early generation of AGB stars.}
{At this stage, the r+s scenario (in which two distincts sources have produced a mixed r/s abundance pattern) cannot be excluded.}

The occurrence of PIEs (hence the i-process) in AGB stars remains very sensitive to the adopted overshoot parametrization, especially with respect to the $f_{\top}$ parameter, as shown in this work. A strong overshoot favors the occurrence of PIEs, even in solar-metallicity AGB stars. 
Constraints on the overshoot parameters (e.g., from multi-dimensional hydrodynamical models) could deliver new insights into the contribution of AGB stars to the heavy element nucleosynthesis, thus offering a step forward in understanding the chemical evolution of the Universe.

\section*{Acknowledgments}
This work was supported by the Fonds de la Recherche Scientifique-FNRS under Grant No IISN 4.4502.19. 
L.S. and S.G. are senior F.R.S-FNRS research associates. A.C. is post-doctorate F.R.S-FNRS fellow.



\bibliographystyle{aa}
\bibliography{astro.bib}


\clearpage
\newpage


\begin{appendix}

\section{Comparison with observations in r/s stars}

\begin{figure*}[h!]
 \begin{minipage}[c]{2\columnwidth}
\includegraphics[width=0.5\columnwidth]{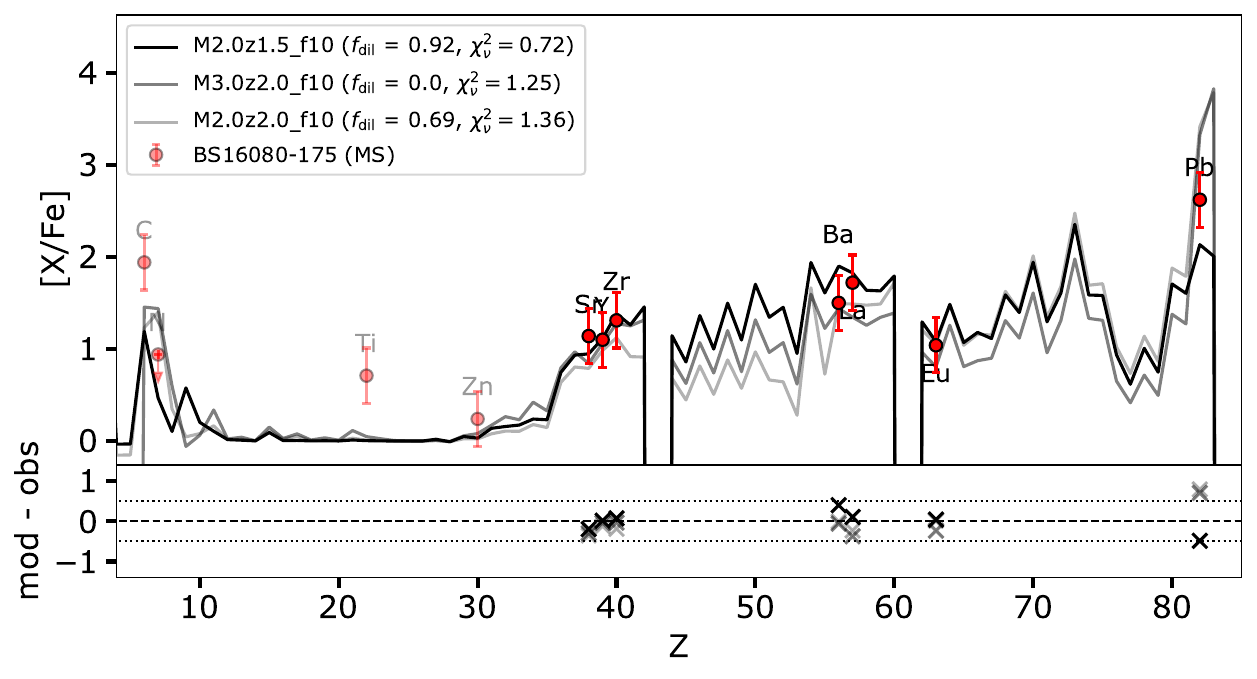}
\includegraphics[width=0.5\columnwidth]{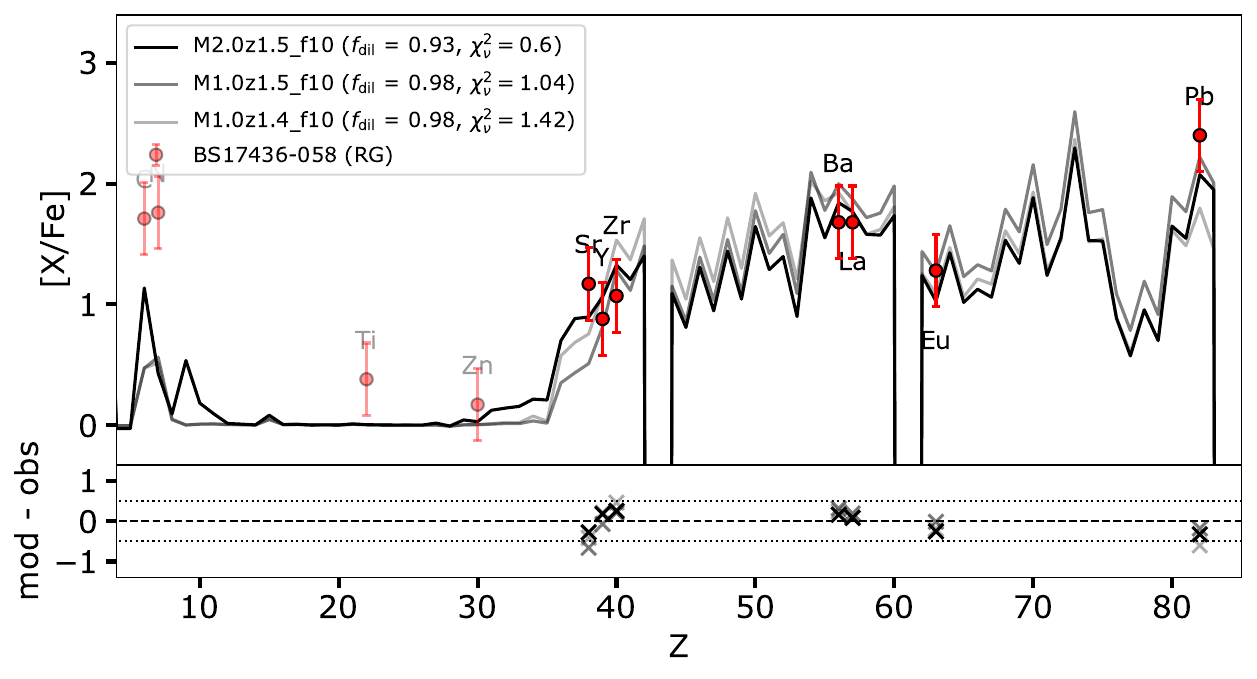}
  \end{minipage}
\begin{minipage}[c]{2\columnwidth}
\includegraphics[width=0.5\columnwidth]{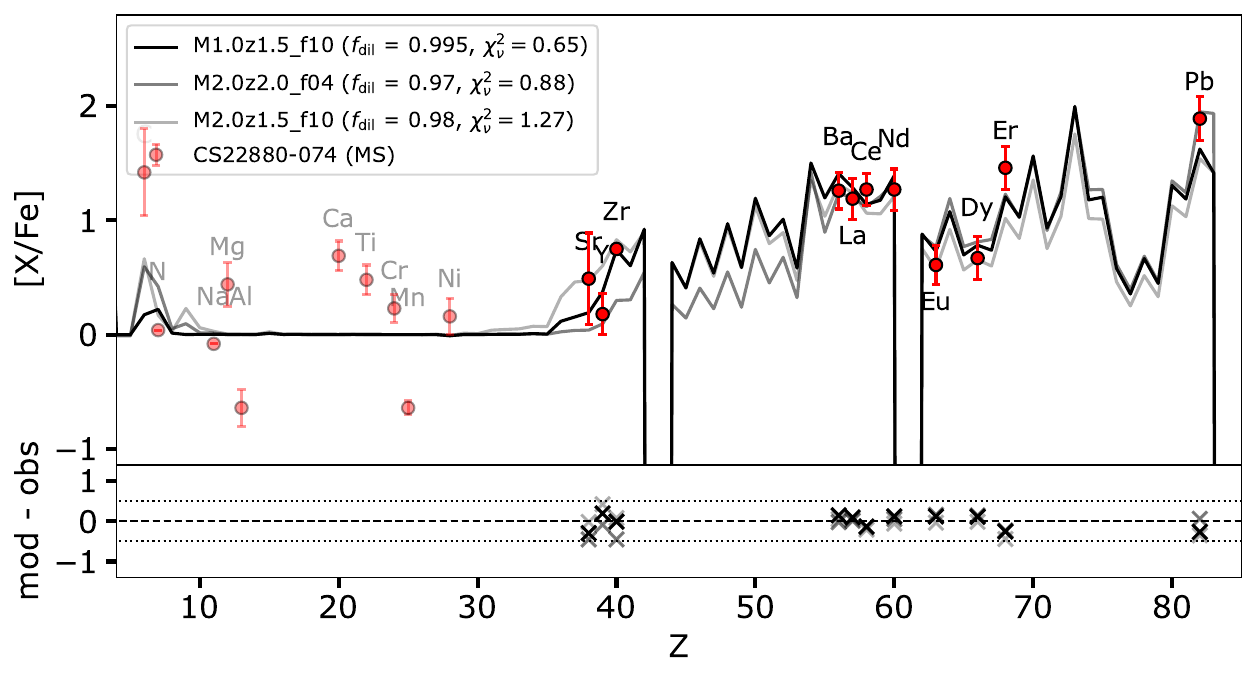}
\includegraphics[width=0.5\columnwidth]{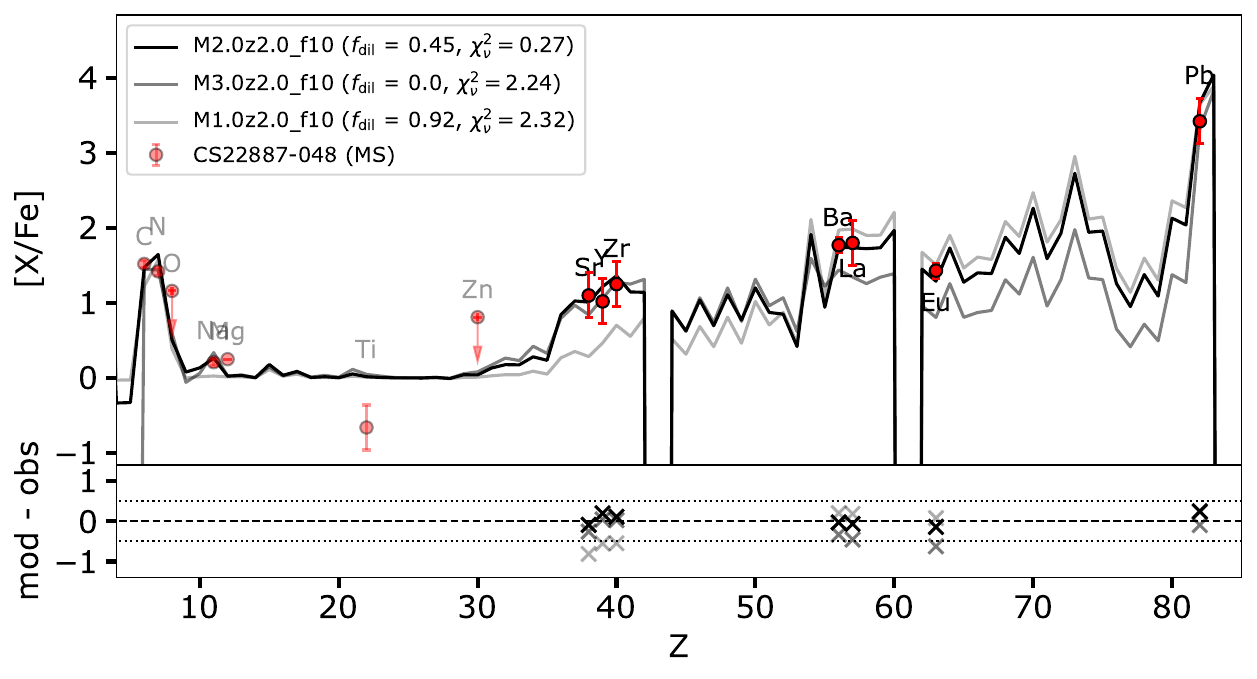}
  \end{minipage}
\begin{minipage}[c]{2\columnwidth}
\includegraphics[width=0.5\columnwidth]{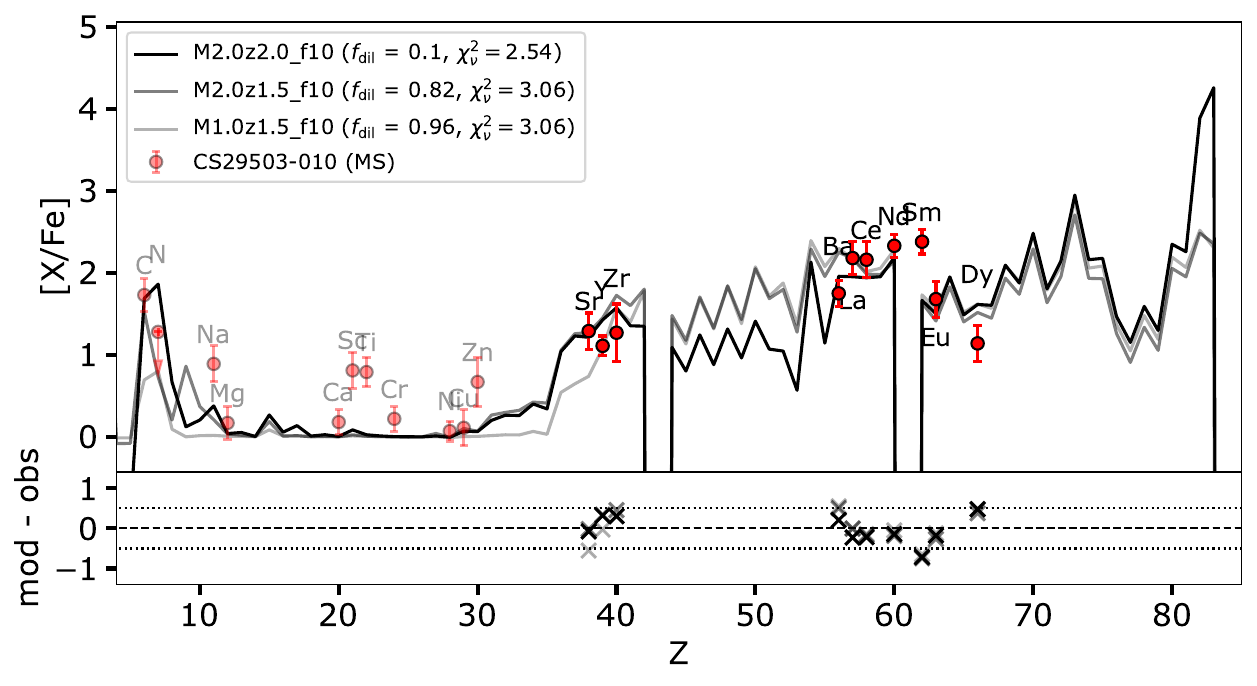}
\includegraphics[width=0.5\columnwidth]{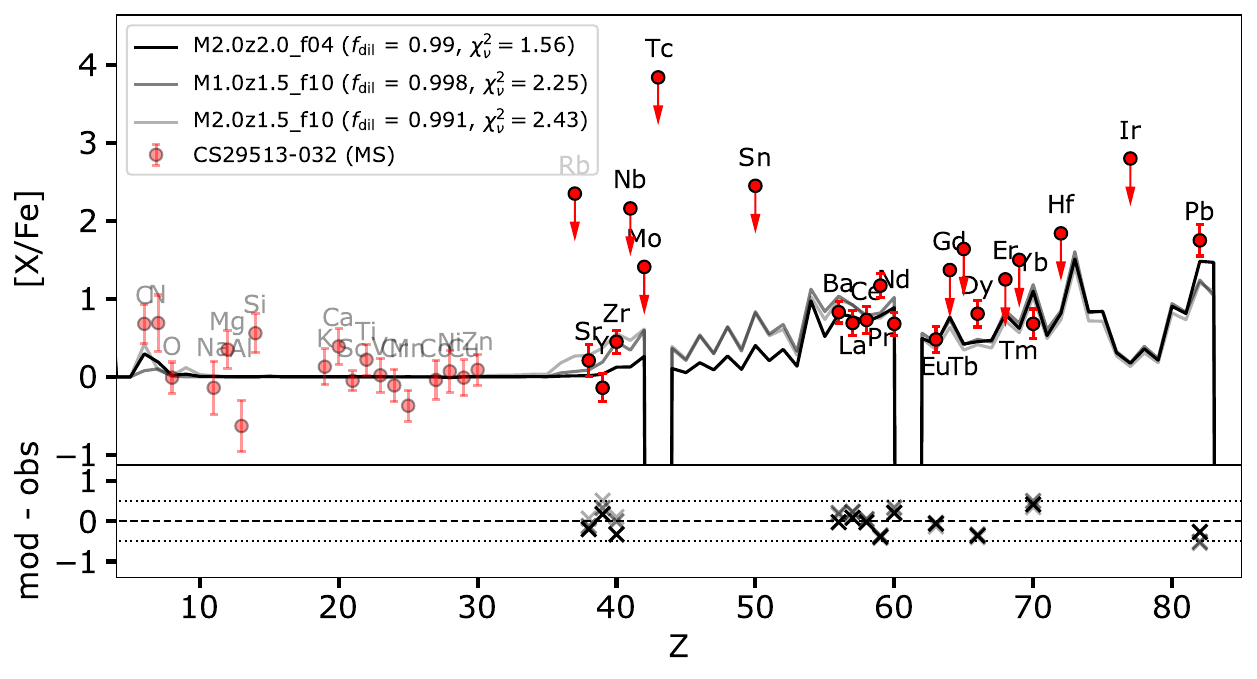}
  \end{minipage}
 \begin{minipage}[c]{2\columnwidth}
\includegraphics[width=0.5\columnwidth]{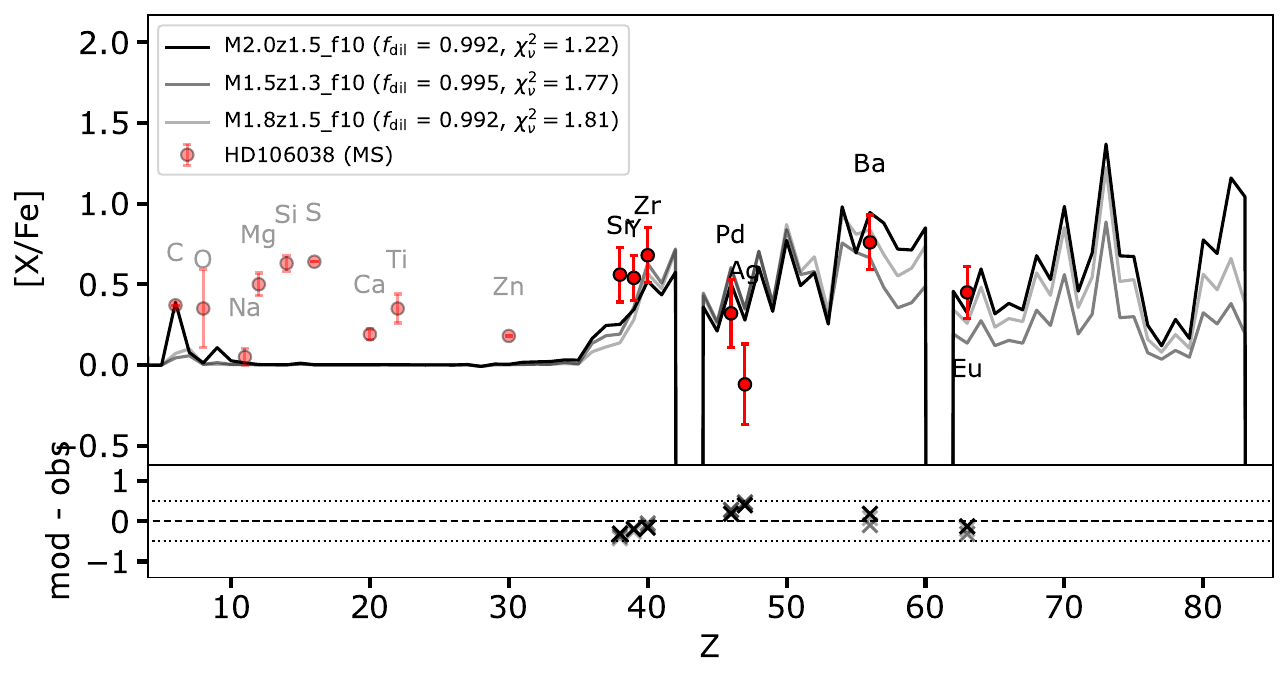}
\includegraphics[width=0.5\columnwidth]{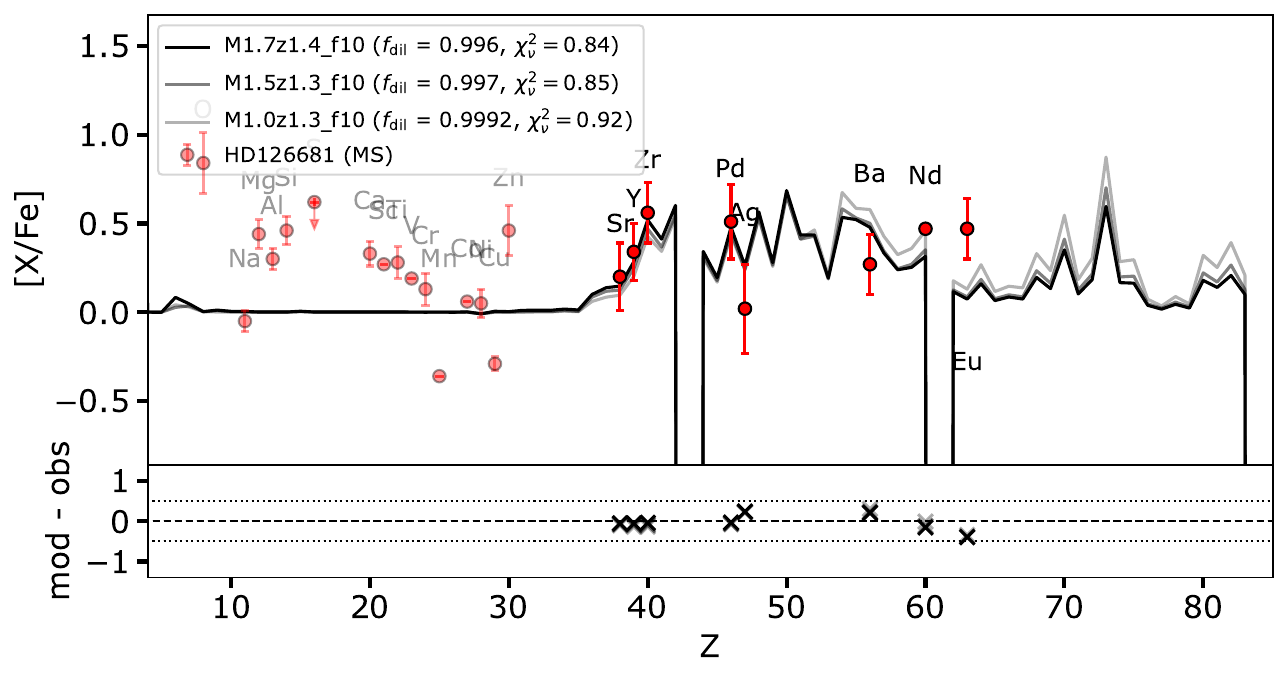}
  \end{minipage}
\caption{Best fits to the sample stars of Table~\ref{table:fits} using the AGB models computed in this work (Table~\ref{table:3}). The upper panels compare the [X/Fe] ratios as a function of the charge number $Z$, while the lower panels gives the deviations between the model and observation. The three best models are shown in black (lowest $\chi_{\nu}^2$), grey (second lowest $\chi_{\nu}^2$), and light grey (third lowest $\chi_{\nu}^2$). The dilution factor $f_{\rm dil}$ (Eq.~\ref{eq:dilf}) and smallest $\chi_{\nu}^2$ value are indicated. 
The abundance data are taken from the SAGA database \citep{suda08}, {complemented with a few recent observations (see Table~\ref{table:fits})}. 
}
\label{fig:fits}
\end{figure*}

Figure~\ref{fig:fits} shows the individual best fits to the possible i-process stars reported in Table~\ref{table:fits}.

\begin{figure*}[h!]
   \ContinuedFloat
\begin{minipage}[c]{2\columnwidth}
\includegraphics[width=0.5\columnwidth]{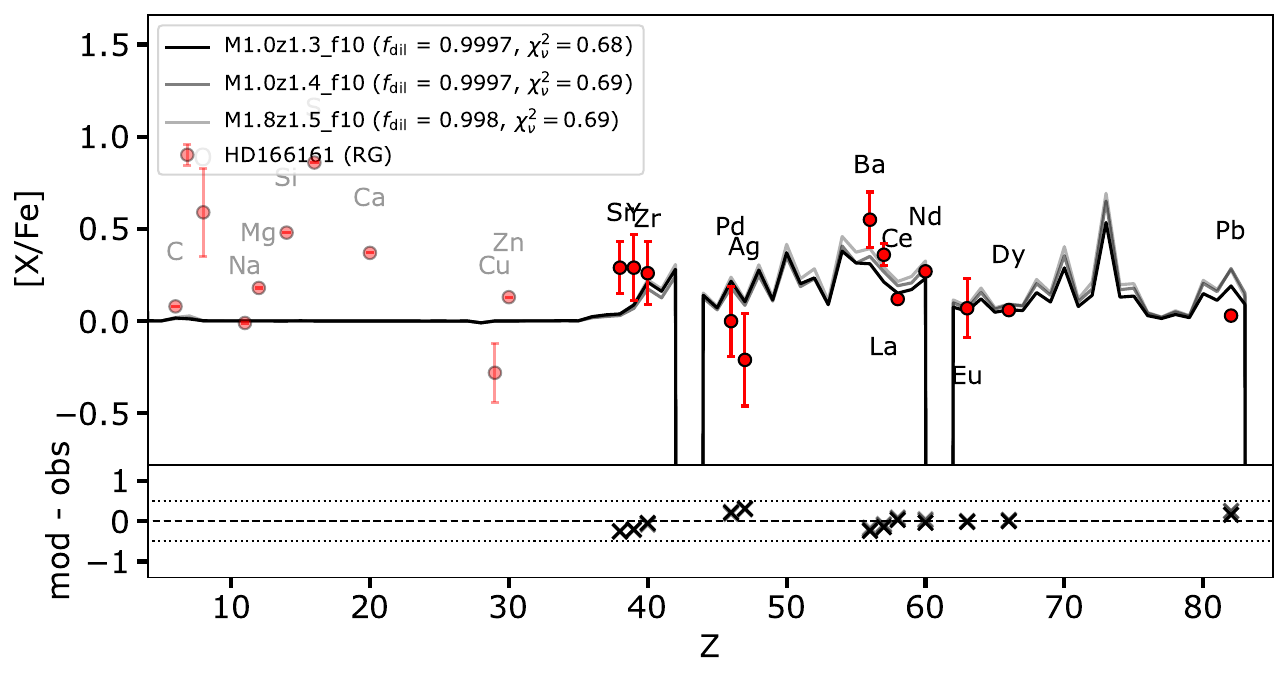}
\includegraphics[width=0.5\columnwidth]{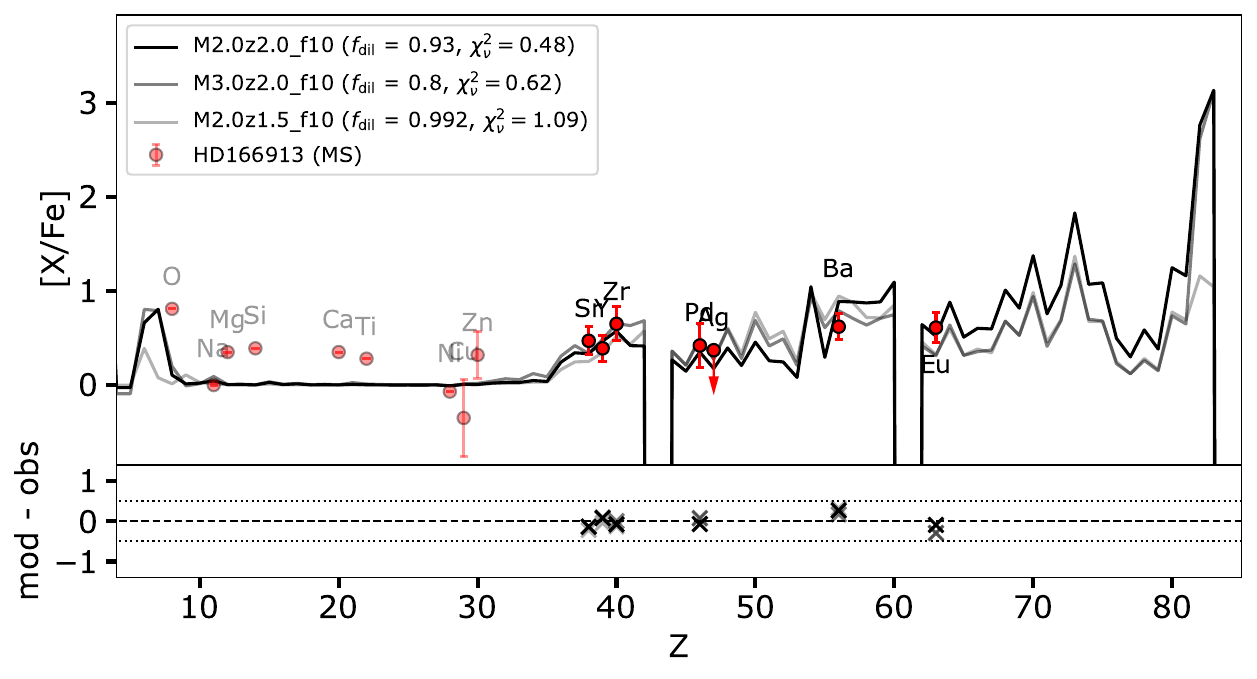}
  \end{minipage}
\begin{minipage}[c]{2\columnwidth}
\includegraphics[width=0.5\columnwidth]{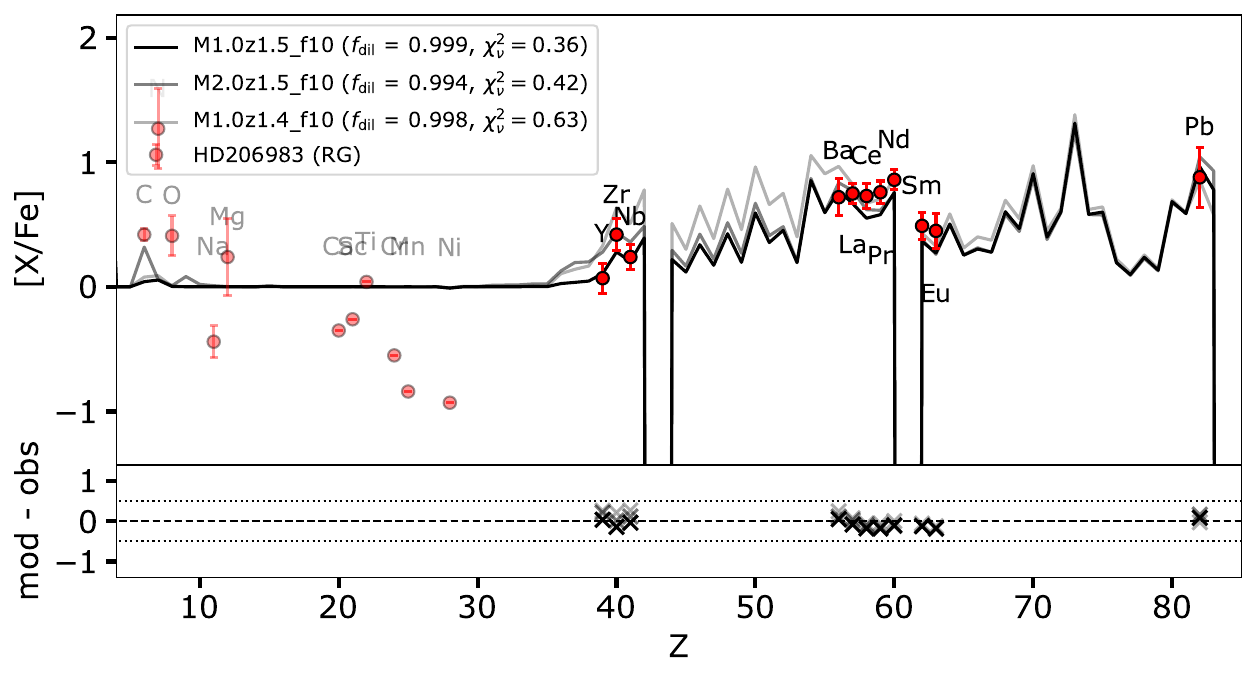}
\includegraphics[width=0.5\columnwidth]{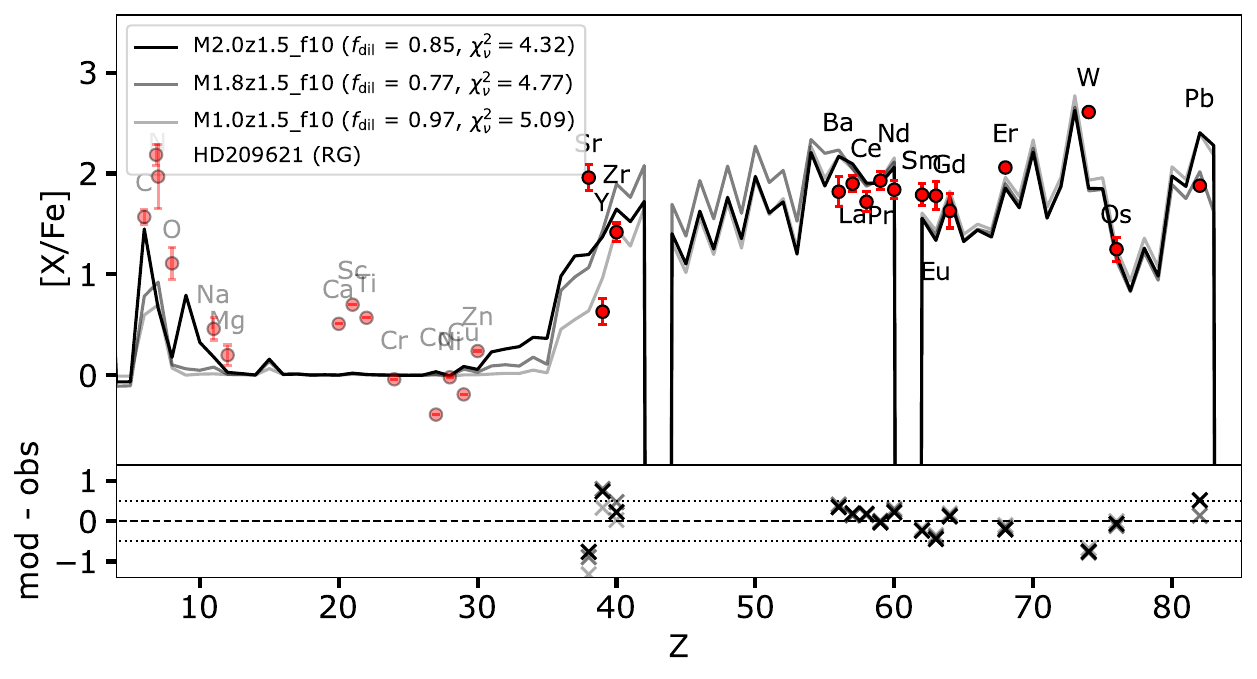}
  \end{minipage}
\begin{minipage}[c]{2\columnwidth}
\includegraphics[width=0.5\columnwidth]{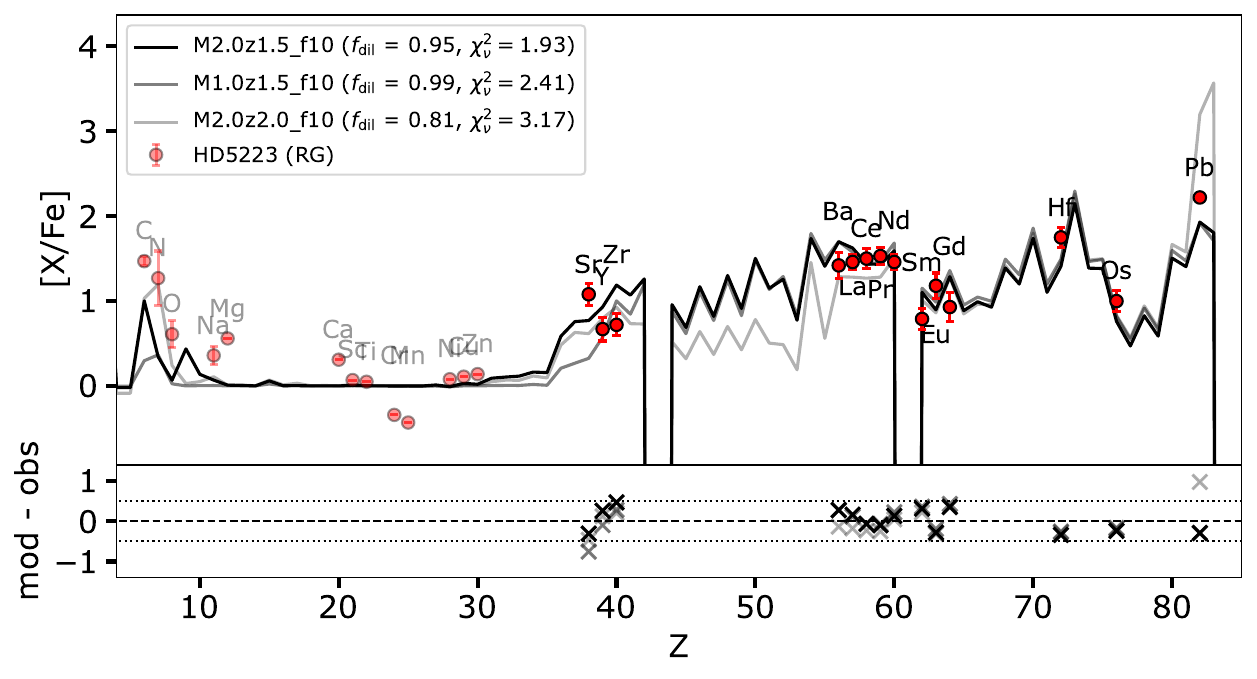}
\includegraphics[width=0.5\columnwidth]{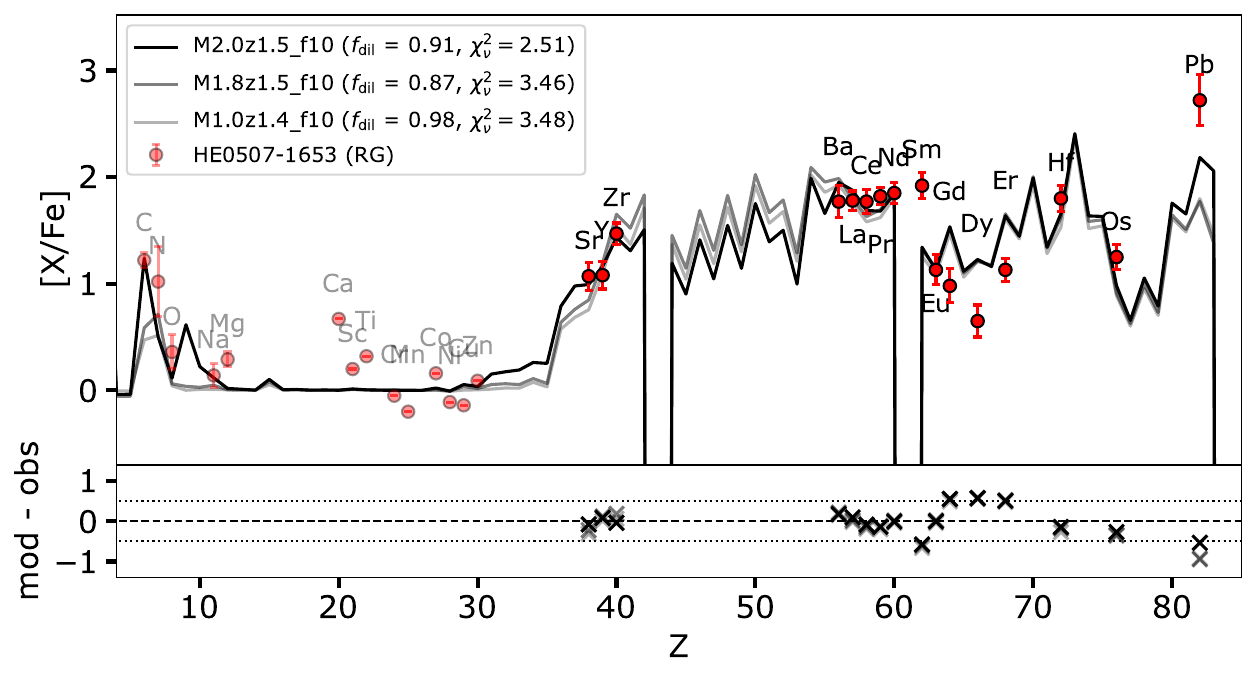}
  \end{minipage}
\begin{minipage}[c]{2\columnwidth}
\includegraphics[width=0.5\columnwidth]{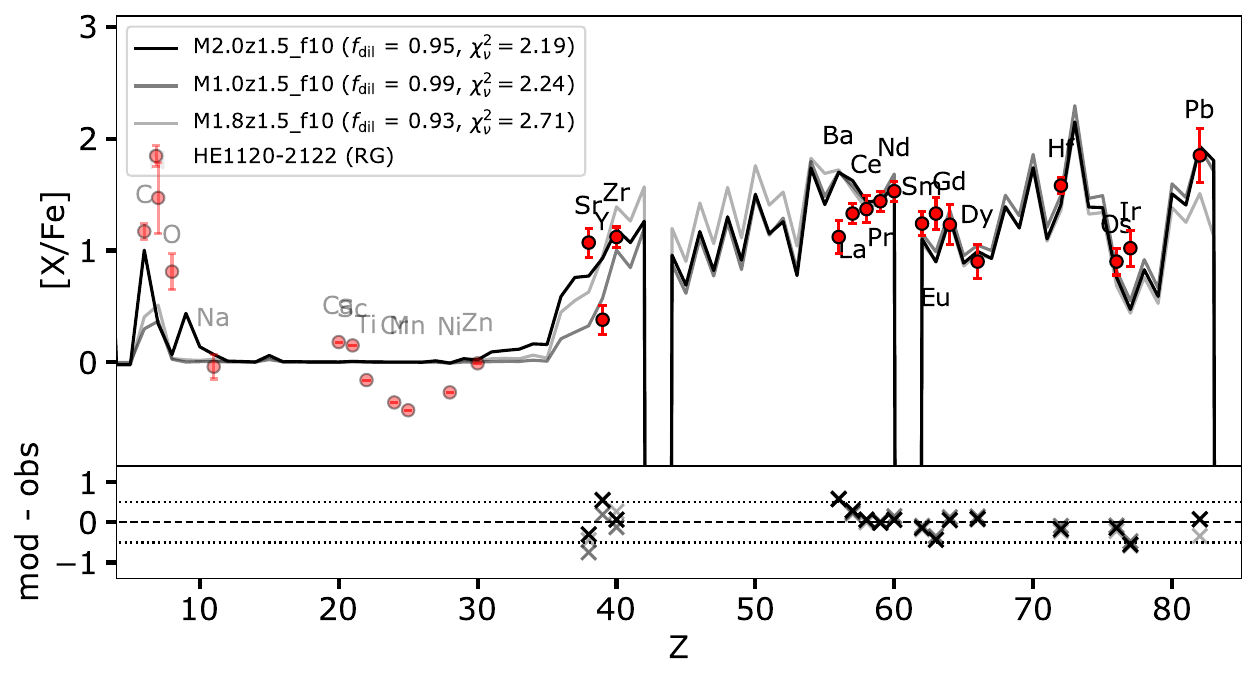}
\includegraphics[width=0.5\columnwidth]{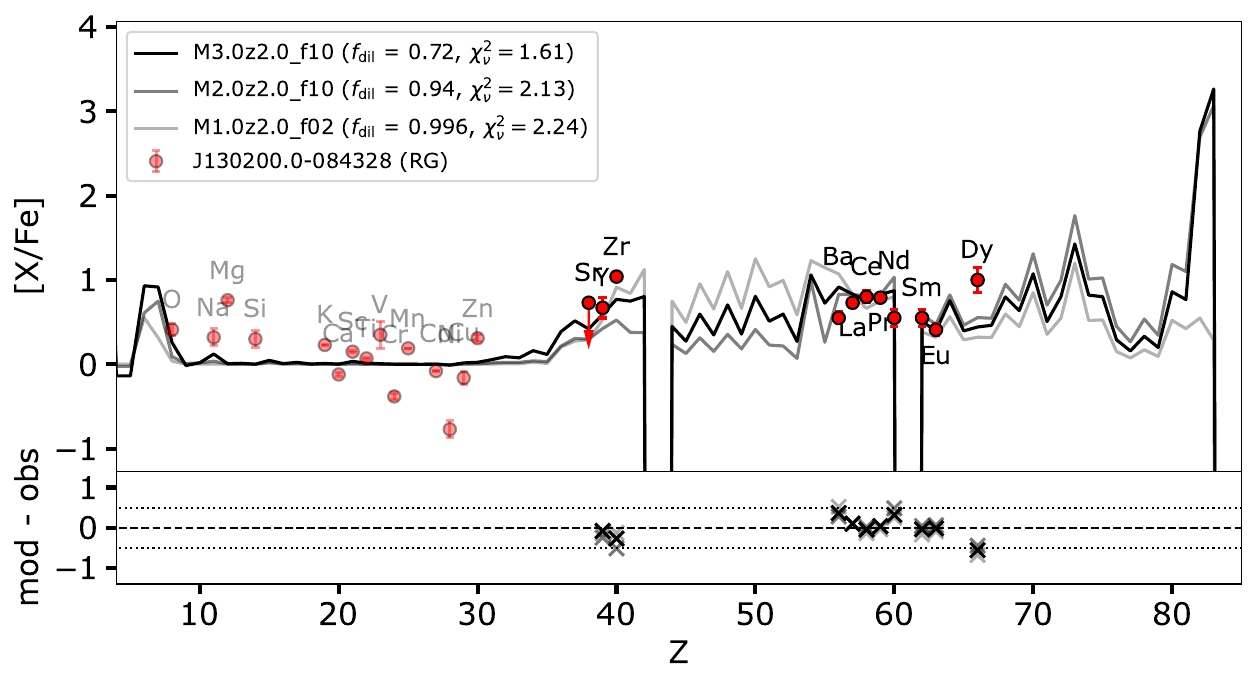}
  \end{minipage}
  \caption{
Continued.
}
\end{figure*}

  \begin{figure*}[h!]
   \ContinuedFloat
\begin{minipage}[c]{2\columnwidth}
\includegraphics[width=0.5\columnwidth]{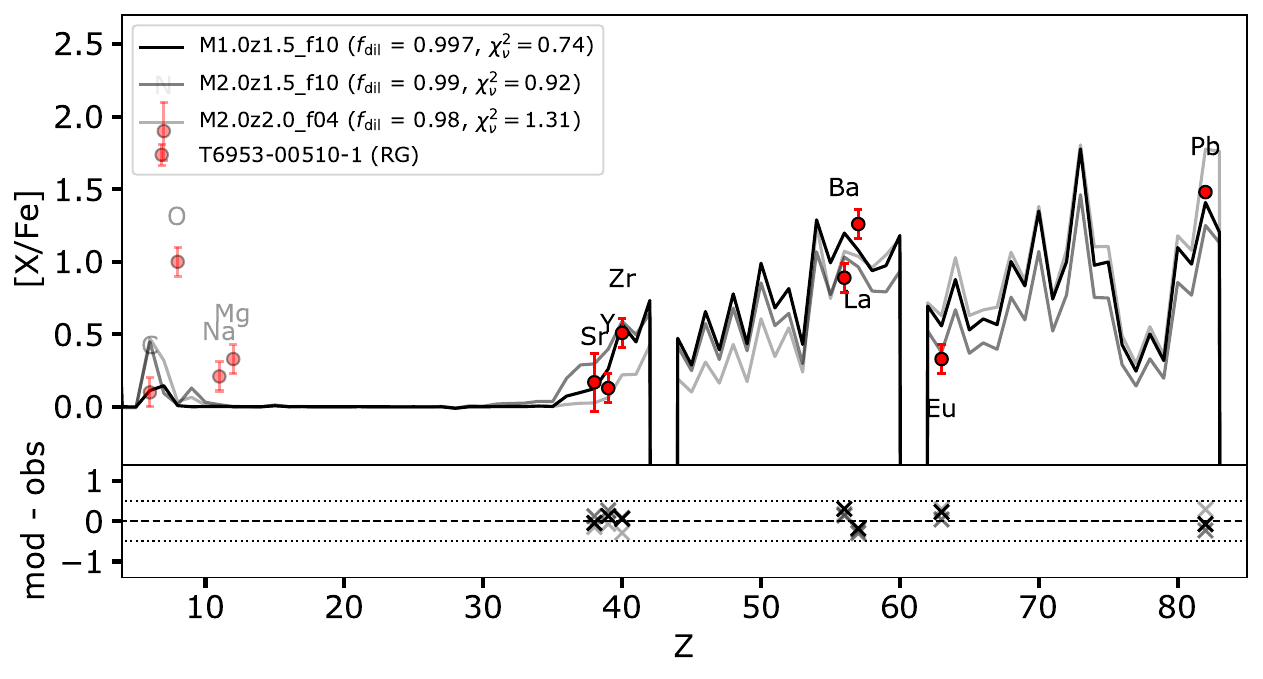}
\includegraphics[width=0.5\columnwidth]{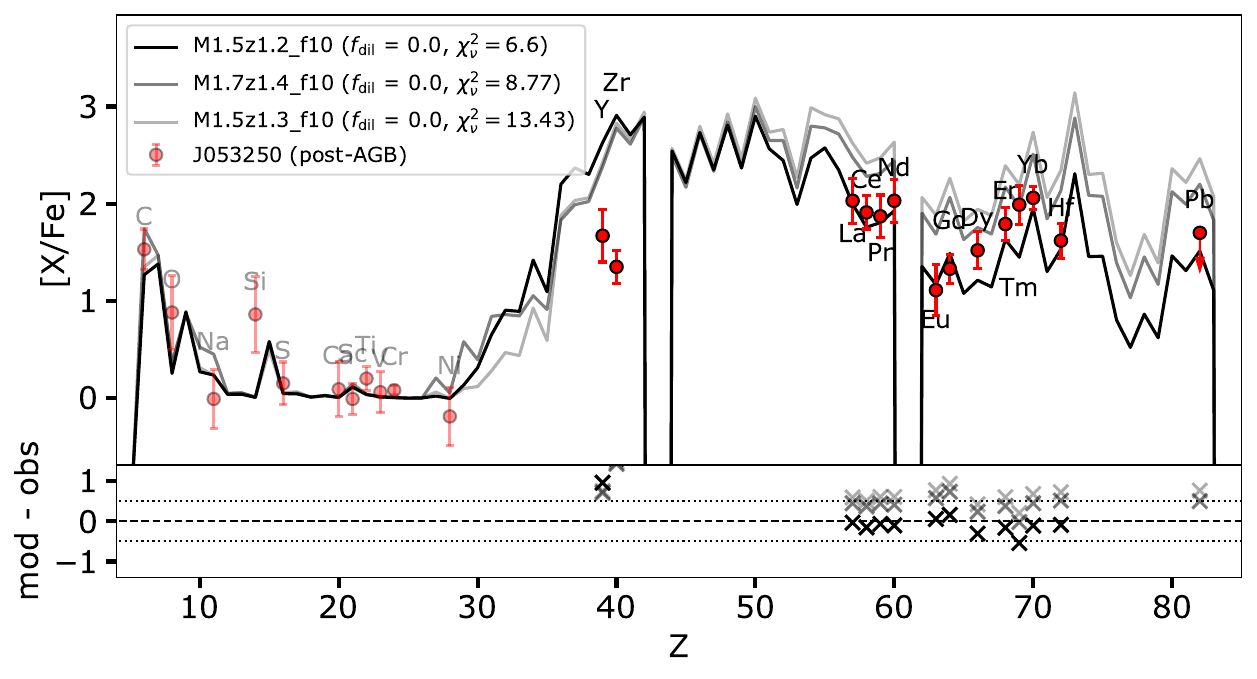}
  \end{minipage}
 \begin{minipage}[c]{2\columnwidth}
\includegraphics[width=0.5\columnwidth]{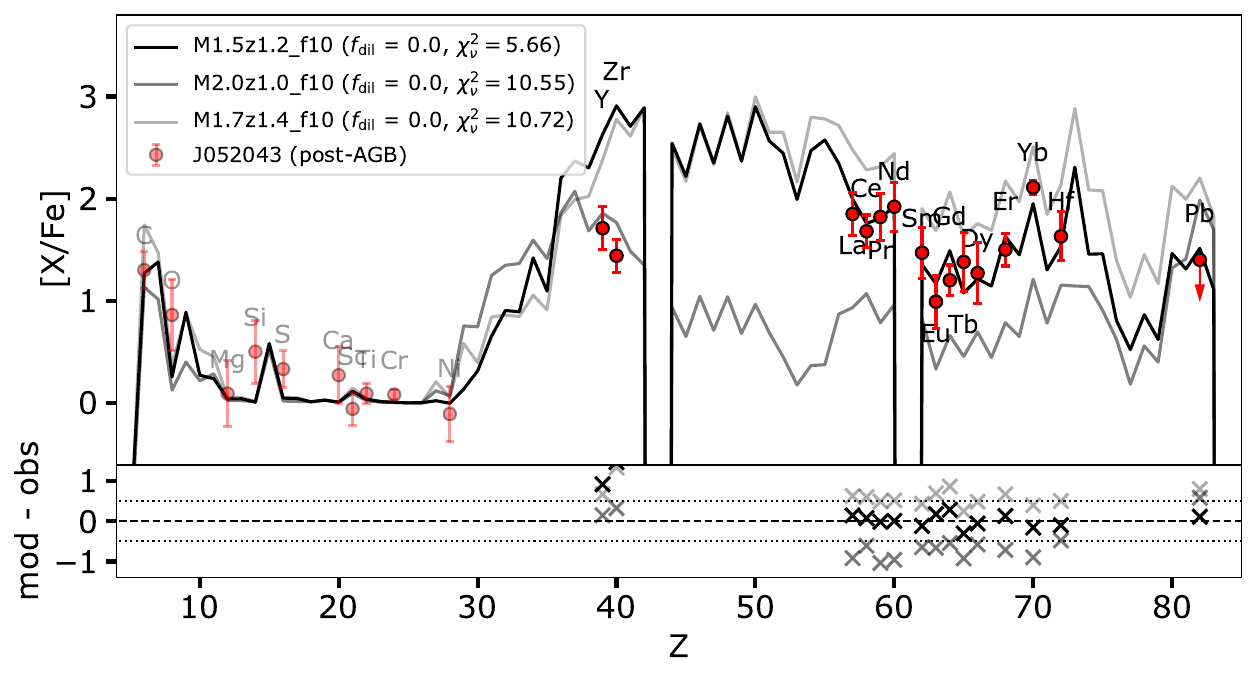}
\includegraphics[width=0.5\columnwidth]{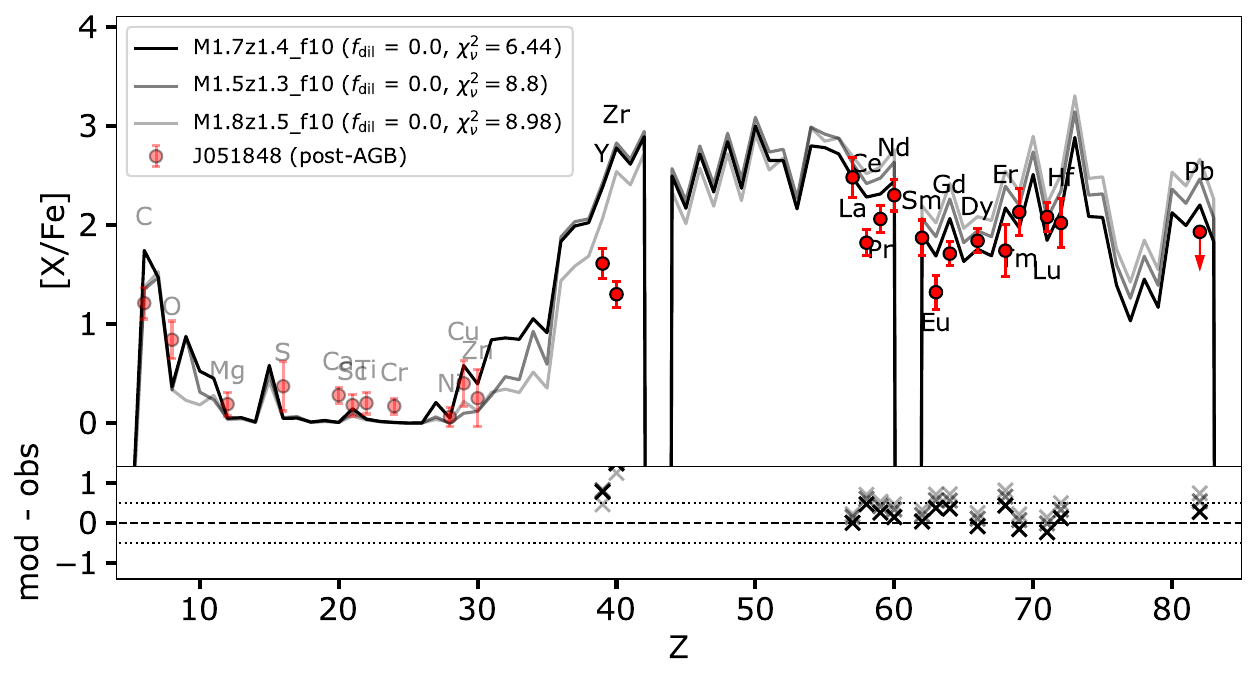}
  \end{minipage}
\caption{
Continued.
}
\end{figure*}

\end{appendix}

\end{document}